\documentclass[9pt, aps, nofootinbib, prd, twocolumn]{revtex4-1}
\usepackage{amsmath}
\usepackage{amssymb}
\usepackage{bm}
\usepackage[usenames,dvipsnames]{color}
\usepackage{dcolumn}
\usepackage{epsfig}
\usepackage{graphicx}
\usepackage{newtxmath, newtxtext}
\usepackage[pagebackref=false, colorlinks=true]{hyperref}
\usepackage{subfigure}
\usepackage{mathrsfs}

\definecolor{reddish}{rgb}{0.7,0.2,0.0}  
\definecolor{blueish}{rgb}{0.1,0.1,1}
\hypersetup{
linkcolor=reddish,         
citecolor=blueish,           
filecolor=blue,              
urlcolor=magenta}       

\begin{document}

\title{Accurate mapping of spherically symmetric black holes in a
  parameterised framework}
\author{Prashant Kocherlakota$^{\text{1}}$}
\author{Luciano Rezzolla$^{\text{1,2}}$}
\affiliation{$^{\text{1}}$Institut f{\"u}r Theoretische Physik,
  Goethe-Universit{\"a}t, Max-von-Laue-Str. 1, 60438 Frankfurt, Germany}

\affiliation{$^{\text{2}}$School of Mathematics, Trinity College, Dublin 2, Ireland}

\begin{abstract}
The Rezzolla-Zhidenko (RZ) framework provides an efficient approach to
characterize spherically symmetric black-hole spacetimes in arbitrary
metric theories of gravity using a small number of variables
\cite{Rezzolla_Zhidenko14}. These variables can be obtained in principle
from near-horizon measurements of various astrophysical processes, thus
potentially enabling efficient tests of both black-hole properties and
the theory of general relativity in the strong-field regime. Here, we
extend this framework to allow for the parametrization of arbitrary
asymptotically-flat, spherically symmetric metrics and introduce the
notion of a 11-dimensional (11D) parametrization space $\Pi$, on which
each solution can be visualised as a curve or surface. An $\mathscr{L}^2$
norm on this space is used to measure the deviation of a particular
compact object solution from the Schwarzschild black-hole solution. We
calculate various observables, related to particle and photon orbits,
within this framework and demonstrate that the relative errors we obtain
are low (about $10^{-6}$). In particular, we obtain the innermost stable
circular orbit (ISCO) frequency, the unstable photon-orbit impact
parameter (shadow radius), the entire orbital angular speed profile for
circular Kepler observers and the entire lensing deflection angle curve
for various types of compact objects, including non-singular and singular
black holes, boson stars and naked singularities, from various theories
of gravity. Finally, we provide in a tabular form the first 11
coefficients of the fourth-order RZ parameterization needed to describe a
variety of commonly used black-hole spacetimes. When comparing with the
first-order RZ parameterization of astrophysical observables such as the
ISCO frequency, the coefficients provided here increase the accuracy of
two orders of magnitude or more.
\end{abstract}

\maketitle

\section{Introduction}
The Dicke-E{\"o}tv{\"o}s experiment established that the trajectories of
freely falling test bodies are independent of their internal structures
and compositions, thereby setting the weak equivalence principle (WEP) on
firm footing. Truly remarkable tests of whether the speed of light is
isotropic and independent of the velocity of the source or not, and tests
of time-dilation, conservation of four-momentum, and the relativistic
laws of kinematics in particle physics experiments have all bolstered our
confidence in the principles of local Lorentz invariance (LLI) and local
positional invariance (LPI) as being fundamental features of any serious
physical theory. Therefore, the Einstein equivalence principle (EEP),
which requires LLI, LPI, and WEP all to hold, is well supported by
experiments to date. For further details, we direct the reader to see the
foundational papers in experimental gravitation \cite{Dicke64,
  Thorne_Will71}, and an excellent modern review can be found in
\cite{Will14}.

Assuming the exact validity of the EEP implies that metric theories of
gravity are the most viable candidates to describe classical gravity, or
possibly theories that are metric apart from very weak or short-range
non-metric couplings (as in string theory) \cite{Dicke64, Thorne_Will71,
  Will14}. Following \cite{Thorne_Will71}, we define a metric theory as
one in which a metric tensor $\boldsymbol{g}$ exists and is necessarily
associated with gravity, and matter and other non-gravitational fields
obey $\boldsymbol{\nabla} \cdot \boldsymbol{T} = 0$, where
$\boldsymbol{\nabla}$ is defined with respect to the metric
$\boldsymbol{g}$, and $\boldsymbol{T}$ is the energy-momentum-stress
tensor for all matter and non-gravitational fields
\cite{Thorne_Will71}. The latter condition on $\boldsymbol{T}$ has the
important consequence that test bodies move on geodesics of
$\boldsymbol{g}$, which is of central importance here \cite{Misner+73}.
Gravitational redshift, bending of light due to spacetime curvature,
frame-dragging effects due to matter currents, and the Shapiro delay are
to be expected in any metric theory of gravity, and one can test
candidate theories quantitatively, in the weak-field limit, for their
agreement with such observables within the parametrized post-Newtonian
(PPN) parametrization scheme proposed in \cite{Will71a, Will71b}, in
terms of 10 variables.

General relativity (GR; \cite{Einstein16}), which is the Ockham's razor
theory of gravity, has withstood all classical weak-field tests to date
successfully \cite{Will14, Collett+18}, and an early success of GR in the
strong-field regime was the prediction of the rate of energy loss due to
gravitational wave radiation in binary pulsar systems
\cite{Taylor_Weisberg82}. More recently, with major large-scale astronomy
missions such as the Laser Interferometric Gravitational wave Observatory
(LIGO) and the Event Horizon Telescope (EHT; \cite{Akiyama+19_1,
  Akiyama+19_2, Akiyama+19_3, Akiyama+19_4, Akiyama+19_5, Akiyama+19_6}),
it is becoming possible to observe astrophysical events that are
dominated by strong-gravity effects. Direct detections of gravitational
waves by LIGO from various compact binary systems \cite{Abbott+16a,
  Abbott+16b} and the recently obtained image of the supermassive compact
object M87$^\star$ by EHT \cite{Akiyama+19_1, Akiyama+19_2, Akiyama+19_3,
  Akiyama+19_4, Akiyama+19_5, Akiyama+19_6} can be interpreted
consistently with the use of the black hole (BH) solutions of GR. The
recent observations by GRAVITY of the gravitational redshift
\cite{Abuter+18} and geodetic orbit-precession \cite{Abuter+20} of the
star S2 near our galaxy's central supermassive compact object Sgr
A$^\star$ are other key successes of GR in strong-gravitational fields.

Various non-BH solutions, such as boson stars and naked singularities,
also exhibit many of the features that BH solutions do, such as the
presence of photon spheres \cite{Olivares+18, Shaikh+18}, and
characterizing observable differences of such ``mimickers" from the BHs
of GR is clearly important. Attempting to address the question of the
validity of the cosmic censorship hypothesis from an observational point
of view is an attractive possibility: while general results like the
Birkhoff theorem \cite{Birkhoff23}, along with various other analytical
\cite{Sasaki_Nakamura90, Dafermos_Rodnianski10, Lucietti_Reall12,
  Duztas_Semiz13, Dafermos+14, Duztas15, Shlapentokh-Rothman15,
  Natario+16, Richartz16} and numerical studies \cite{Sasaki_Nakamura82,
  Miller_Motta89, Yo+02, Baiotti+05, Nathanail+17} lend weight to our
expectation that BHs can, in fact, occur frequently, or equivalently that
they do form generically as endstates of continual gravitational
collapse, despite significant effort \cite{Christodoulou84,
  Christodoulou86, Shapiro_Teukolsky92, Choptuik93, Christodoulou94,
  Christodoulou99b, Harada+02, Crisford_Santos17} we have not been able
to rule out the formation of naked singularities in GR.

Of course, the very presence of spacetime singularities, which are
locations of arbitrarily large curvatures, in the various solutions of GR
is a long standing weakness of the theory. Their existence is assuredly
generic \cite{Penrose65, Penrose69, Penrose79, Hawking_Ellis73}, and
their formation is independent of whether or not they sheathed behind
event horizons (see, e.g., \cite{Lemaitre33_Tolman34_Bondi47,
  Eardley_Smarr79}). Therefore, it is useful to study observables
associated with regular solutions for BH-like compact objects, both
within GR and in alternative theories of gravity, to check whether they
are consistent with recent strong-gravity measurements of, e.g., the
M87$^\star$ shadow size recently obtained by the EHT, and to explore
whether they are better models for compact astrophysical objects.

Since the number of models for compact objects offered by various
candidate theories of gravity (sometimes when coupled to other fields) is
large, when attempting to test the theory of general relativity using
strong-field observables, it is imperative that we have a unified
theory-agnostic framework ready that characterizes arbitrary solutions
(BHs and non-BHs) efficiently, i.e., with as few parameters as
possible. Towards this end, we extend here the framework presented in
\cite{Rezzolla_Zhidenko14}, which can be used to test properties of
asymptotically-flat, spherically symmetric BH solutions from arbitrary
metric theories of gravity, to include non-BH solutions as well. Our
parametrization framework uses 11 parameters and we are able to obtain
approximate values for the metrics and various observables, for a variety
of compact objects, at typical relative errors of $10^{-6}$. The
observables we chose to study here are the orbital angular speeds of test
bodies moving on circular geodesics, the impact parameter of photons on
unstable circular geodesics (shadow radius), and the angle of deflection
due to gravitational lensing; a study of these observables is important
when considering the construction of images of compact objects from
general-relativistic magnetohydrodynamic (GRMHD) simulations. We also
report here the deviations of these observables from their corresponding
values for the Schwarzschild BH, for easy comparison.

Finally, since EEP may only hold approximately, i.e., since it could be 
violated in the strong-field regime (see, e.g., \cite{Damour96, Damour12}), 
it is imperative that the framework we use here to set up strong-field tests 
of theories of gravity be able to characterize BH solutions from theories that 
break, e.g., LLI (as in Einstein-aether theories \cite{Eling+04}) or even 
non-metric theories in which, e.g., the electromagnetic Lagrangian 
is modified to allow for non-linear interactions \cite{Ayon-Beato_Garcia99}. Therefore, 
the models for compact objects we consider here are: BHs from (a) GR that 
are either singular \cite{Schwarzschild16, Reissner16_Nordstrom18} or
non-singular \cite{Bardeen68, Hayward06, Held+19, Frolov16}, (b)
Einstein-aether theory \cite{Berglund+12}, (c) string theory
\cite{Kazakov_Solodhukin94, Gibbons_Maeda88, Garfinkle+92, Garcia+95,
  Casadio+02}, and (d) GR coupled to non-linear electrodynamics
\cite{Bronnikov01, Yajima_Tamaki01}. Additionally, we also consider
spacetimes of regular mini boson stars \cite{Olivares+18} and naked
singularities \cite{Janis+68} in GR. We argue that since the level of errors in
approximating their exact observables is sufficiently low, it is possible
to distinguish between these objects extremely well whenever their exact
variables differ, within the present framework.

The outline of the paper is as follows. In
Sec. \ref{sec:eRZ_Parametrization} we discuss a unified framework to
parametrize and implement strong- and weak-field tests of arbitrary
spherically symmetric metrics in arbitrary metric theories of gravity
(and some that are non-metric, as mentioned above). We note that this is
a smooth extension of the Rezzolla-Zhidenko parametrization scheme
presented in \cite{Rezzolla_Zhidenko14}. In
Sec. \ref{sec:Observables_eRZ}, we outline how various observables
related to causal geodesics may be computed within this parametrization
scheme. In Sec. \ref{sec:Spacetimes_eRZ}, we introduce the notion of a
11D parametrization space $\Pi$ on which every metric solution can be
uniquely visualised, and provide brief descriptions of the various
compact objects under consideration here. We also demonstrate the
efficiency of our framework in obtaining the metric functions (up to two
derivatives) across the entire region of interest. For example, for BHs,
we are able to approximate their entire exterior geometry to a maximum
relative error that is typically less than $10^{-6}$. We also show how
all of the associated observables considered here are recovered at
similar error levels. Sec. \ref{sec:Discussion_Summary} presents a
summary of our results and discusses various advantages of this
framework. Our results have considerable overlap with the analysis for
BHs presented in \cite{Konoplya_Zhidenko20}, and we briefly compare the
two sets of results in Sec. \ref{sec:Discussion_Summary}.

\section{An efficient parametrization framework for spherically symmetric spacetimes}
\label{sec:eRZ_Parametrization}

The Rezzolla-Zhidenko (RZ) framework of parameterizing
asymptotically-flat, spherically symmetric BH spacetimes in arbitrary
metric theories of gravity \cite{Rezzolla_Zhidenko14} effectively
rewrites a portion of the metric functions in terms of
continued-fractions over a conformal radial coordinate. A smooth
extension of this scheme was proposed in \cite{Konoplya+16} to tackle the
problem of parameterizing the broader class of asymptotically-flat,
axially-symmetric BH spacetimes, when the metric is expressed in
Boyer-Lindquist-like coordinates ($t, r, \theta, \phi$). The existence of
the two Killing vector fields $\partial_t$ and $\partial_\phi$ ensures
that the four free metric functions depend only on $r$ and $\theta$, and
in the Konoplya-Rezzolla-Zhidenko (KRZ) framework \cite{Konoplya+16}, a
double-expansion in these variables is employed to parametrize them. In
particular, a Taylor expansion in $y = \cos{\theta}$ and a mixed
Taylor-Pad\'e expansion in terms of a conformal radial coordinate $x$,
similar to the one used here, efficiently parametrizes the exterior
horizon geometry. We direct the reader towards \cite{Younsi+16} for a
demonstration of the efficiency of the KRZ scheme in parameterizing
various well known stationary BH metrics and their associated shadow
curves. 

Restricting to spherically symmetric spacetimes, we now discuss an
extension of the RZ scheme that allows for arbitrary asymptotically-flat,
static spacetimes, including non-BH ones, to also be similarly
characterised.

The line element outside a spherically symmetric configuration of 
matter can generally be expressed in arbitrary spherical-polar coordinates $(t, \rho,
\theta, \phi)$ as
\begin{equation} \label{eq:General_Static_Metric}
\text{d}s^2 = - f(\rho)\text{d}t^2 + g(\rho)\text{d}\rho^2 + h(\rho)\text{d}\Omega_2^2\,,
\end{equation}
where $\text{d}\Omega_2^2$ is the standard line element of a
two-sphere. Since the aim of the current parametrization scheme is to
compare metrics across arbitrary metric theories of gravity, it is useful
to re-express them in a standardized form, in the same set of 
``areal-radial, polar'' coordinates $(t, r, \theta, \phi)$ as,
\begin{equation} \label{eq:General_Static_Metric_RZ}
\text{d}s^2 = - N^2(r)\text{d}t^2 + \frac{B^2(r)}{N^2(r)}\text{d}r^2 + r^2 \text{d}\Omega_2^2\,.
\end{equation}
This radial coordinate $r$ cleanly determines the proper-area of
two-spheres $\mathcal{A}$ in the spacetime as, $\mathcal{A}= 4\pi r^2$.
The desired coordinate transformation $\rho \rightarrow r$ to achieve
this change in form can be obtained by solving for $\rho(r)$ from,
\begin{equation}  \label{eq:RZ_Gauge}
h(\rho) = r^2\,,
\end{equation}
with the other metric functions then being given as, $N^2(r) = f(\rho(r))$ and
$B^2(r) = f(\rho(r))g(\rho(r))\left(\partial_r\rho(r)\right)^2$, where
$\partial_r$ represents a derivative with respect to $r$.

One can then compactify the radial coordinate by introducing an interior
cutoff for it at $r = r_0 > 0$ and defining a conformal radial coordinate
$x$ as\footnote{Note that $x$ is not a conformally-\textit{flat} 
  coordinate. See Appendix \ref{app:Other_Useful_Coordinates} for a 
  discussion on how $x$ is related to the conformally-flat ``tortoise'' 
  coordinate $r_*$.},
\begin{equation} \label{eq:Conformal_Coordinate}
x(r) = 1 - \frac{r_0}{r}\,,
\end{equation}
and the coordinate patch we will be interested in characterizing here is
$r_0 \! \leq \! r \! < \! \infty$. Clearly, $x(r  =  r_0) = 0$ and as
$r \rightarrow \infty, x(r) \rightarrow 1$. Therefore, characterizing the
metric functions $N^2(x)$ and $B^2(x)$ over the range $0 \! \leq x \! <
\! 1$ is equivalent to fully characterizing the spacetime over this
radial range. It is useful to keep in mind the nature of this scale;
i.e., a radial range $r_0 \! < \!  r \! < \!  2 r_0$ takes up almost half
of the range of the conformal coordinate $0 \!  < \!  x \!  < \!
.5$. Also, the range $10^4 r_0 \!  < \! r \! < \!  10^6 r_0$ is packed
into $1-10^{-4} \! < \! x \! < \! 1-10^{-6}$.

When a metric \eqref{eq:General_Static_Metric_RZ} describes the geometry
outside a BH, a natural choice for $r_0$ exists, namely the location of
its event horizon, since one is typically interested in studying features
of its exterior geometry. Indeed, this will be our choice
here\footnote{To be precise, when various types of horizons for a BH
  solution do not match (e.g., Einstein-{aether} BHs \cite{Berglund+12}),
  we will always choose $r_0$ to correspond to the outermost Killing
  horizon, which is the location of the outermost zero of the null
  expansion, and is given by the outermost root of $g_{rr}^{-1}$, i.e.,
  $N^2(r_0) = 0$. }.

Similarly, if one is interested in studying the exterior geometry of a
star, one can set $r_0$ to correspond to the location of its surface. In
the case of a spacetime containing no such natural interior boundary,
like that of a boson star or a naked singularity, one can set $r_0$
freely to a finite non-zero value. Since the central objective of the
present study is to study differences of metric functions and observables
associated with various compact objects from the Schwarzschild BH in
particular, a convenient choice for $r_0$ here, for such objects, is $r_0
= 2M$, where $M$ is the Arnowitt-Deser-Misner (ADM; \cite{Arnowitt+62})
mass of the spacetime. Since we will be considering asymptotically-flat
spacetimes exclusively here, identifying $M$ is typically possible. This
also reduces the number of requisite parameters from 12 to 11, as we
will see below.
 
As noted above, a fundamental necessity to be able to constrain
deviations from GR is a unified theory-agnostic framework that
characterizes both the strong- and weak-gravitational field regimes of
arbitrary solutions efficiently. Since, by construction, the RZ
parametrization scheme handles both these regimes simultaneously and
effectively for spherically symmetric BH solutions
\cite{Rezzolla_Zhidenko14}, a natural choice is to extend it to include
non-BH solutions. This is achieved by modifying the auxiliary function
$A(x)$ used in \cite{Rezzolla_Zhidenko14} as,
\begin{equation} \label{eq:Ax_eRZ}
N^2(x) = n_0+ A(x)x\,,
\end{equation}
where $n_0  =  N^2(r  =  r_0)$. In particular, when the metric
\eqref{eq:General_Static_Metric} describes a BH spacetime we have $n_0 \!
= \! 0$, and this definition for $A(x)$ reduces to the one used in
equation (4) of \cite{Rezzolla_Zhidenko14}. We have essentially modified
the ``inner''-boundary condition on the 1D box $0 \! \leq x \! < \! 1$ for
the $g_{tt}$-metric function.

For the Killing vector $\partial_t$ to remain timelike
($\bm{g}(\partial_t\cdot\partial_t) = g_{tt} < 0$) outside the outermost
Killing horizon, clearly, we require,
\begin{equation} \label{eq:BHs_A_non-negative}
0 < A(x), \ \ \text{for}\ \ 0 \! < x \!  < \! 1\,.
\end{equation}

If a non-BH spacetime admits a Killing surface at some location (perhaps
in aether theories), one must set $r_0$ to correspond to that
location. The non-BH spacetimes considered here do not admit such Killing
surfaces\footnote{The absence of a Killing horizon implies stationary
  timelike Killing observers with four-velocities $u \propto \partial_t +
  \Omega \partial_\phi$ exist all the way to the centre of the spacetime;
  these move on circular orbits. In the region where $N > r N_{,r} > 0$,
  equatorial circular geodesics ($\nabla_u u = 0$) exist [see
    Eq. \eqref{eq:Kepler_Frequency} below]. For the non-BH spacetimes
  considered here, such equatorial Kepler observers can exist all the way
  to the centre.  }, and we will set $r_0 = 2M$ for them in
Sec. \ref{sec:Spacetimes_eRZ}.

Equation \eqref{eq:BHs_A_non-negative} implies a continued-fraction
approximation for $A(x)$ is already a salient possibility. However, to
facilitate an easy comparison with the PPN form of the metric
\cite{Will71a, Will14}, we first write out the asymptotic Taylor
expansions (to the first few orders) of the metric functions $A$ and $B$,
and introduce two new auxiliary functions $\tilde{A}$ and $\tilde{B}$ as,
\begin{align}
A(x) =&\ 1 - n_0 - \epsilon(1 - x) + (a_0 - \epsilon)(1 - x)^2 \label{eq:A_x}\\
& \quad + \tilde{A}(x)(1 - x)^3\,, \nonumber\\
B(x) =&\ 1 + b_0(1 - x) + \tilde{B}(x)(1 - x)^2\,. \label{eq:B_x}
\end{align}
In the above, we have also introduced three new constants $\epsilon,
a_0$, and $b_0$, which, along with $n_0$, can be used to test whether the
spacetime in question satisfies the PPN constraints that arise from
weak-field tests of gravity, as we will see in
Sec. \ref{sec:PPN_Constraints}. This redefinition (\ref{eq:A_x},
\ref{eq:B_x}) of the auxiliary functions has the consequence that these
tilded auxiliary functions, $\tilde{A}$ and $\tilde{B}$, do not influence
the values of the PPN parameters of the spacetime.

Thus far, we have roughly performed Taylor expansions of the metric
functions when rewritten in terms of $x$ (a variable that behaves as
$1/r$) about $x = 0$ in equation \eqref{eq:Ax_eRZ} and $x = 1$ in
equations (\ref{eq:A_x}, \ref{eq:B_x}). At the core of the efficiency of
the current parametrization scheme is the choice to characterize
$\tilde{A}$ and $\tilde{B}$ as Pad\'e approximants in the form of
continued-fractions as,
\begin{equation} \label{eq:tAtB_Pade}
\tilde{A}(x) = \cfrac{a_1}{1 + \cfrac{a_2 x}{1 + \cfrac{a_3 x}{1 + \cdots}}}\,,\ \ \tilde{B}(x) = \cfrac{b_1}{1 + \cfrac{b_2 x}{1 + \cfrac{b_3 x}{1 + \cdots}}}\,.
\end{equation}
Therefore, the set of PPN coefficients $n_0, \epsilon, a_0$ and $b_0$,
along with the Pad\'e expansion coefficients, $a_i$ and $b_i$ ($i > 0$),
completely characterize arbitrary spherically symmetric spacetimes in
arbitrary metric theories of gravity.

By definition, these coefficients $a_i\ (i > 0)$) can be obtained by
Taylor-expanding the continued-fractions in equation
\eqref{eq:tAtB_Pade}, and matching the Pad\'e expansion coefficients
order-by-order with the Taylor expansion coefficients for $\tilde{A}(x)$,
which we write formally as,
\begin{equation} \label{eq:Taylor_Expansion_tA}
\tilde{A}(x) = \sum_{i=0}^{\infty}\tilde{a}_{i+1} x^i\,.
\end{equation}
We show below the first few Pad\'e coefficients of a function
$\tilde{A}(x)$ in terms of its Taylor coefficients,
\begin{align}
a_1 =&\ \tilde{a}_1\,,\ \
a_2 =\ -\frac{\tilde{a}_2}{\tilde{a}_1}\,,\ \
a_3 =\ -\frac{\left(\tilde{a}_3\tilde{a}_1 - \tilde{a}_2^2\right)}{\tilde{a}_2\tilde{a}_1}\,, \\
a_4 =&\ -\frac{\left(\tilde{a}_4\tilde{a}_2 - \tilde{a}_3^2\right)\tilde{a}_1}{(\tilde{a}_3\tilde{a}_1 - \tilde{a}_2^2)\tilde{a}_2}\,, \nonumber \\
a_5 =&\ -\frac{\left(\tilde{a}_5\left(\tilde{a}_3\tilde{a}_1 - \tilde{a}_2^2\right) - \tilde{a}_4^2\tilde{a}_1 + 2\tilde{a}_4\tilde{a}_3\tilde{a}_2 - \tilde{a}_3^3\right)\tilde{a}_2}{\left(\tilde{a}_4\tilde{a}_2 - \tilde{a}_3^2\right)\left(\tilde{a}_3\tilde{a}_1 - \tilde{a}_2^2\right)}\,,\nonumber \\
a_6 =& -\frac{\left(\tilde{a}_6\left(\tilde{a}_4\tilde{a}_2 - \tilde{a}_3^2\right) - \tilde{a}_5^2\tilde{a}_2 + 2\tilde{a}_5\tilde{a}_4\tilde{a}_3 - \tilde{a}_4^3\right)\left(\tilde{a}_3\tilde{a}_1 - \tilde{a}_2^2\right)}{\left(\tilde{a}_5\left(\tilde{a}_3\tilde{a}_1 - \tilde{a}_2^2\right) - \tilde{a}_4^2\tilde{a}_1 + 2\tilde{a}_4\tilde{a}_3\tilde{a}_2 - \tilde{a}_3^3\right)\left(\tilde{a}_4\tilde{a}_2 - \tilde{a}_3^2\right)}\,, \nonumber 
\end{align}
to demonstrate that the map between two sets expansion coefficients for
the same function is non-linear. The reader may have observed that the
dependence of either type of coefficient on the other of the same order
is linear.

Henceforth, by an $n^{\text{th}}$-order approximation we will mean that
we have truncated the Pad\'e approximants by setting $a_{i > n} = 0$. The
power of the present parametrization scheme is primarily due to well
known property of the rapidity of the order-on-order convergence of
Pad\'e approximants \eqref{eq:tAtB_Pade} to the exact value for a
multitude of functions \cite{Baker_Graves-Morris96, Bender_Orszag99}, as
compared to other approximation schemes such as Taylor expansions
\eqref{eq:Taylor_Expansion_tA}, for example. That this well known
efficiency of Pad\'e approximants is carried into the RZ parametrization
has been demonstrated for the Einstein-dilaton spacetime
\cite{Rezzolla_Zhidenko14}, where the rate of convergence of
order-on-order truncated Pad\'e approximants was contrasted against the
order-on-order truncated Taylor approximations of the Johannsen-Psaltis
parametrization scheme in the appendix of \cite{Rezzolla_Zhidenko14}. In
appendix \ref{app:Carson_Yagi} below, we conduct a similar convergence
test against a recently proposed Taylor expansion-based parametrization
scheme \cite{Carson_Yagi20} for the Bardeen-BH metric. See also appendix
\ref{app:Additional_Tests_eRZ}.

As we will see below in Sec. \ref{sec:Spacetimes_eRZ}, already at the
fourth-order ($a_5 = b_5 = 0$), we are able to recover both metric
functions for various spacetimes with a maximum relative error of about
$10^{-6}$ over the entire range $0 \leq x <1$, and not just close to the
boundaries of the spacetime. In particular, for BH spacetimes, the
relative errors are typically far lower at this order. In fact, it has
recently been argued that already at the second order one can recover the
unstable photon-orbit radius, the orbital angular frequency of the
innermost Kepler observer, and the quasi-normal frequency spectrum for
scalar perturbations to the desired accuracy for BH solutions that are
not close to extremality \cite{Konoplya_Zhidenko20}.

For (metric) functions that are fractions of two polynomials, the
associated continued-fractions only have a finite (and typically small)
number of coefficients $a_i$, i.e., for some $n$, all $a_{i>n} = 0$, and
the $n^{\text{th}}$-order Pad\'e approximant converges exactly to the
exact function (see for example the case of the Einstein-aether 1 BH in
Sec. \ref{sec:Spacetimes_eRZ}). However, as can be seen from the cases of
the Bronnikov BH and the Janis-Newman-Winicour naked singularity below,
even for non-polynomial functions, the Pad\'e approximant still converges
fairly rapidly.

Note that it is not always possible to set a particular coefficient
$a_{n+1}$ to zero in order to obtain the $n^{\text{th}}$-order truncated
Pad\'e approximant. One such instance is easily seen when $a_n < - 1$; In
this case, setting $a_{n+1} = 0$ creates a pole at $0 \leq x = -1/a_n <
1$ for the $n^{\text{th}}$-order approximant.  To get around such an
obstacle, following the discussion in Sec. IV of \cite{Konoplya+16}, we
may simply set $a_{n+2} = 0, a_{n+1} = 1$ and obtain then the
approximation at the $n^{\text{th}}$-order.

It is now reasonable to ask how small a particular Pad{\'e} coefficient
$a_n$ needs to be in order for higher-order coefficients to be
neglected. One finds that such zeroes, $a_n \rightarrow 0$, are typically
associated with poles at the next order, $a_{n+1} \rightarrow \pm\infty$,
and it is clear that the combined effect of this zero-pole pair is to
send $1 + {a_n x}/{(1+a_{n+1} x)} \rightarrow 1$. Therefore, we argue
that when $|a_n|$ is appropriately small, we can set $a_{i \geq n} = 0$.

We turn finally to the inner-boundary behaviour of the metric functions
in this parametrization scheme. The Taylor expansion of the metric
functions near $x=0$ is given as,
\begin{align}
N^2 =&\ n_0 + \left(1 - n_0 - 2\epsilon + a_0 + a_1\right)x\\ 
+&\ \left(3\epsilon - 2a_0 -3a_1 -a_1 a_2\right)x^2 + O\left(x^3\right)\,, \nonumber \\
\frac{B^2}{N^2} =&\ \frac{n_0}{\left(1+b_0+b_1\right)^2}\\
+&\ \left(\frac{\left(1 - n_0 - 2\epsilon + a_0 + a_1\right)}{\left(1+b_0+b_1\right)^2} + \frac{2n_0\left(b_0 + 2b_1 + b_1 b_2\right)}{\left(1+b_0+b_1\right)^3}\right)x \nonumber \\
+&\ O\left(x^2\right)\,. \nonumber 
\end{align}
In the case of a BH spacetime ($n_0 = 0$), $x$ measures the distance from
the horizon and therefore the above expressions capture the near-horizon
geometry of a BH.

We end by noting that the condition given in equation
\eqref{eq:BHs_A_non-negative} has the effect of setting non-trivial
constraints on the allowed ranges of the expansion parameters ${\epsilon,
  a_i}$ for BH solutions, i.e., when working at a particular order $n$,
the expansion parameters $\epsilon, a_{i \leq n}$ cannot be freely
chosen. This is of considerable importance when employing this
parametrization scheme to set up tests by solving inverse problems.

\section{Characterizing observables in the parametrization scheme}
\label{sec:Observables_eRZ}
In this section, we outline how various observables associated with
spherically symmetric metrics can be obtained within the current
parametrization scheme. In particular, we discuss how the PPN
parameters, the orbital angular frequency of Kepler observers, and
deflection of light due to gravitational lensing can be calculated within
this framework. We show also the calculation for the impact parameter of
photons on unstable circular geodesics for completeness
\cite{Rezzolla_Zhidenko14}. In addition to these observables, the method
to obtain the quasi-normal frequencies associated with scalar
perturbations of spherically symmetric spacetimes within this
parametrization scheme can also be found in
\cite{Rezzolla_Zhidenko14}. We find it useful to note here that of the
observables considered here, only the gravitational lensing deflection angle depends on
the metric function $B$. When two spacetimes have identical
$N^2$-functions, this observable can be used to distinguish between the
two spacetimes (see for example the instances of the Hayward and Modified
Hayward BHs in Sec. \ref{sec:Spacetimes_eRZ}).

\subsection{Testing PPN constraints} \label{sec:PPN_Constraints}
The metric functions corresponding to a generic asymptotically-flat
spacetime can be expanded around asymptotic infinity, $x = 1$, and
expressed as,
\begin{align} \label{eq:PPN_Compactified_Form}
N^2 =&\ 1 - \frac{2M}{r_0}(1-x) + (\beta - \gamma)\frac{2M^2}{r_0^2}(1-x)^2 + O\left((1-x)^3\right)\,, \nonumber\\
\frac{B^2}{N^2} =&\ 1 + \gamma\frac{2M}{r_0}(1 - x) + O\left((1-x)^2\right)\,,
\end{align}
where $\beta$ and $\gamma$ are parameters that can be obtained from the
fall-off features of the metric functions. For metric theories, $\beta$
and $\gamma$ are called Parametrized Post-Newtonian (PPN) parameters, and
measure respectively, e.g., the agreement of their predictions for the
perihelion shift of Mercury and the time delay or light deflection due to
the Sun; these satisfy \cite{Will14},
\begin{equation} \label{eq:PPN_Constraints}
|\beta - 1| \lesssim 2.3 \times 10^{-4}\,,\ \ |\gamma -1| \lesssim 2.3 \times 10^{-5}\,.
\end{equation}

The Taylor expansions of the metric functions around the exterior
boundary of the spacetime can be obtained as,
\begin{align}
N^2 =&\ 1 - (1 - n_0 + \epsilon)(1-x) + a_0(1-x)^2 + O\left((1-x)^3\right)\,, \nonumber \\
\frac{B^2}{N^2} =&\ 1 + (1 - n_0 + \epsilon + 2b_0)(1-x) + O\left((1-x)^2\right)\,, \label{eq:PPN_Form_RZ}
\end{align}
and on comparing equations \eqref{eq:PPN_Compactified_Form} and
\eqref{eq:PPN_Form_RZ}, we can identify that,
\begin{align}
\epsilon = \frac{2M}{r_0} - \left(1 - n_0\right)\,,\ \ a_0 = \frac{2M^2}{r_0^2}(\beta - \gamma)\,,\ \ b_0 = \frac{M}{r_0}(\gamma - 1)\,.
\end{align}
Therefore, the PPN constraints \eqref{eq:PPN_Constraints} then
straightforwardly translate into constraints on the four constants
introduced above $n_0, \epsilon, a_0$, and $b_0$, as,
\begin{align}  \label{eq:PPN_RZ_Param}
\mathcal{P}_1 =&\ \left|\frac{2a_0}{(1 - n_0 + \epsilon)^2} + \frac{2b_0}{(1 - n_0 + \epsilon)}\right| \lesssim 2.3 \times 10^{-4}\,, \nonumber \\
\mathcal{P}_2 =&\ \left|\frac{2b_0}{(1 - n_0 + \epsilon)}\right| \lesssim 2.3 \times 10^{-5}\,.
\end{align}
Note that a spacetime with vanishing zeroth-order parameters, $a_0$ and
$b_0$, straightaway satisfies PPN constraints. Furthermore, it is also
clear that the functions $\tilde{A}$ and $\tilde{B}$ do not contribute in
any capacity towards asymptotic PPN constraints.

For non-BH spacetimes ($n_0 \neq 0$), if one sets $r_0 = 2M$, then
$\epsilon = n_0$, and these constraints simplify to $\mathcal{P}_1 =
|2a_0 + 2b_0|$ and $\mathcal{P}_2 = |2b_0|$ respectively.
Notice that for BH spacetimes, since $n_0 = 0$, the parameter $\epsilon$
must satisfy $\epsilon > -1$ for the horizon to exist ($r_0 > 0$).

We also use the PPN constraints above \eqref{eq:PPN_RZ_Param} for all of
the BH solutions coming from the non-metric theories of gravity used here
since (a) for the dilaton BHs, despite the non-metric coupling of the
electrodynamics (ED) Lagrangian, photons still move along null geodesics
of the metric \eqref{eq:General_Static_Metric_RZ}\footnote{The dilaton
  gravity Lagrangian used here violates WEP in general due to a varying
  fine-structure constant (see, e.g., \cite{Magueijo03}), but not LLI or
  LPI.}, and (b) for the non-linear ED BHs, the Lagrangian reduces to
Einstein-Hilbert-Maxwell in the weak-field limit
\cite{Ayon-Beato_Garcia99, Bronnikov01, Yajima_Tamaki01}. We think it
useful to mention here also that these BHs have the same asymptotic
behaviour as the BHs of GR (up to the relevant orders for PPN
considerations). The Einstein-aether BHs considered here have $\beta =
\gamma = 1$ (see, e.g., \cite{Jacobson07, Jacobson08}). Note that we have
made the rather strong assumption that even though these theories might
not satisfy the Birkhoff theorem these constraints are satisfied by
astrophysical BHs.

\subsection{Photon and particle orbits}
Central to the comparison of images of compact objects that the EHT will
obtain, such as Sgr A$^\star$, against GRMHD simulations is the study of
the flow of matter in accretion disks near such objects, and of the
motion of photons in the associated spacetime. As a first approximation,
if the motion of accreting matter is modelled as being circularly
freely-falling, then a study of timelike stable circular geodesics
becomes important. The radial in-fall speed of matter on such orbits is
negligible compared to the speed at which it rotates around the compact
object. In general, circular Kepler geodesics do not extend all the way
into the black hole or up to the surface of a non-BH compact object, and
there exists an innermost stable circular orbit (ISCO) at some radius $r
 =  r_{_{\text{ISCO}}}$. Matter below this point $r \! < \!
r_{_{\text{ISCO}}}$ is pulled onto the compact object considerably more
quickly. Therefore, local features of the flow of matter differ
significantly depending where on the matter is relative to the ISCO, and
the angular speed of matter at this location $\Omega_{_{\text{ISCO}}}$
sets a dynamical free-fall timescale and constitutes an important
observable of the compact object. Since this matter is typically a hot
plasma, it emits radiation which is lensed by the gravity of the compact
object before it reaches asymptotic observers present on earth for
example. Some of these photons are also trapped by the compact object,
depending on whether or not it possesses a photon sphere, which can be
characterised by the impact parameter of the unstable circular photon
orbit $\xi_{\text{ps}}$. Essentially, (radially in-going) photons with
impact parameter $\xi \! < \! \xi_{\text{ps}}$ are captured by the
central object, and are on unstable orbits, as we will see below. Therefore, the
union of the direction of all unstable null geodesics, from the point of
view of an asymptotic observer, in a spacetime geometry constitutes
its shadow region, and whose boundary is characterised by the photon
sphere \cite{Hioki_Maeda09}.

In this section, we will show how the Kepler orbital angular frequency
profile $\Omega_{\text{K}}(r)$, its ISCO value $\Omega_{_{\text{ISCO}}}
\equiv \Omega_{\text{K}}(r_{_{\text{ISCO}}})$, the light deflection angle
due to gravitational lensing $\Delta\phi_{\text{GL}}$, and the impact
parameter of the photon on a circular unstable geodesic $\xi_{\text{ps}}$
can be obtained within the present parametrization scheme. Towards this
end, we begin with a brief discussion on circular causal geodesics, with
particular focus on unstable null and stable timelike ones. Since we are
concerned with spherically symmetric spacetimes, a discussion of circular
geodesics in the equatorial plane suffices \cite{Hioki_Maeda09}.

The Lagrangian describing geodesic motion in a static spacetime
\eqref{eq:General_Static_Metric} is given by
\begin{equation} \label{eq:Geodesics_Static_Spacetimes}
2\mathcal{L} = - N^2(r)\dot{t}^2+\frac{B^2(r)}{N^2(r)}\dot{r}^2 + r^2\dot{\theta}^2 + r^2\sin^2\theta \dot{\phi}^2\,,
\end{equation}
where the overdot represents a derivative with respect to the affine parameter. Since the Lagrangian is independent of $t$ and $\phi$, one obtains two constants of the motion as,
\begin{equation}
p_t := \frac{\partial\mathcal{L}}{\partial\dot{t}}=-N^2\dot{t}=-E\,,\ \ 
p_\phi := \frac{\partial\mathcal{L}}{\partial\dot{\phi}}=r^2\sin^2\theta \dot{\phi}=L\,,
\end{equation}
where $E$ and $L$ are, respectively, the energy and angular momentum of
the observer. We can rewrite equation
(\ref{eq:Geodesics_Static_Spacetimes}) for geodesics restricted to the
equatorial plane ($\theta = \pi/2, \dot{\theta} = 0$) as,
\begin{equation} \label{eq:Equatorial_Null_Geodesic_Equation}
\frac{B^2}{N^2}\dot{r}^2+ \left(\frac{L^2}{r^2} - \frac{E^2}{N^2} -2\mathcal{L}\right) = 0\,,
\end{equation}
where $2\mathcal{L}  =  0$ for null geodesics and $2\mathcal{L}  =  -1$ for timelike geodesics. Let us define, for convenience, effective potentials for equatorial null ($V$) and timelike ($\tilde{V}$) observers as,
\begin{align} \label{eq:V_eff}
V :=&\ E^2\left(\frac{\xi^2}{r^2} - \frac{1}{N^2}\right)\,, \\
\tilde{V} :=&\ \tilde{E}^2\left(\frac{\tilde{\xi}^2}{r^2} - \frac{1}{N^2} + \frac{1}{\tilde{E}^2}\right)\,,
\end{align}
where in the above we have introduced the impact parameter of a null
observer as $\xi = L/E$ and analogously also the impact parameter
$\tilde{\xi}$ of a timelike observer.

Equatorial circular null geodesics satisfy $\dot{r} = 0$ and $\ddot{r} =
0$, or equivalently $V = 0$ and $\partial_r V = 0$. The stability of a
circular null geodesic is governed by the sign of $\partial_r^2 V$ ($-$
implies unstable). The expressions for the first and second derivatives
of the effective potential are provided below for later use,
\begin{align} \label{eq:Eff_Potential_Derivatives}
\frac{\partial_r V}{E^2} =&\ -2\left(\frac{\xi^2}{r^3} - \frac{\partial_r  N}{N^3}\right)\,, \\ 
\frac{\partial_r^2 V}{E^2} =&\ 6\left(\frac{\xi^2}{r^4} - \frac{(\partial_r N)^2}{N^4} + \frac{\partial_r^2 N}{3 N^3}\right)\,. \nonumber
\end{align}
Stability of circular timelike geodesics can be similarly determined and
the relevant expressions for $\partial_r \tilde{V}$ and $\partial_r^2
\tilde{V}$ for them can be obtained simply by replacing all of the
quantities in equation \eqref{eq:Eff_Potential_Derivatives} with their
tilded counterparts.

\subsubsection{Photon sphere impact factor}
As discussed above, equatorial circular null geodesics satisfy,
\begin{equation}
0 = \frac{\xi^2}{r^2} - \frac{1}{N^2}\,, \ \ 
0 = \frac{\xi^2}{r^3} - \frac{\partial_r N}{N^3}\,. 
\end{equation}
Equivalently, their radii $r  =  r_{\text{c}}$ can be found by solving,
\begin{equation} \label{eq:Photon_Sphere_Equation}
r - \frac{N(r)}{\partial_r N(r)} = 0\,.
\end{equation}
If $\partial_r^2 V(r = r_{\text{c}}) \! < \! 0$, the spacetime has an
unstable circular null geodesic at that location, and a stable circular
null geodesic otherwise. Of these locations, that which corresponds to
the absolute maximum of the null geodesic potential $V$ marks the
boundary of the shadow\footnote{It is to be noted that if there is no
  unstable circular null geodesic that corresponds to the location of the
  global maximum of the effective potential $V$, then such spacetimes do
  not cast shadows. Tangibly, one can imagine a spacetime that satisfies
  $\lim_{r\rightarrow 0} V(r) = \infty$, such as the
  Reissner-Nordstr{\"o}m naked singularity spacetime, over a certain
  range of specific charge.}.

We will denote this location by $r_{\text{ps}}$. The corresponding impact
parameter $\xi_{\text{ps}}$ of a photon on such an orbit is given as,
\begin{equation} \label{eq:xi_Photon_Sphere}
\xi_{\text{ps}} = \frac{r_{\text{ps}}}{N(r_{\text{ps}})}\,.
\end{equation}
While the photon sphere marks the boundary of the shadow region of a
spacetime, when viewing the compact object from asymptotic infinity, due
to gravitational lensing, we see it to be of size $\xi_{\text{ps}}$
\cite{Hioki_Maeda09}, which the EHT has observed. In terms of the
conformal radial coordinate introduced above, $x  =  1 - r_0/r$, we
can now find the location of all allowed circular null geodesics by
finding the solution $x  =  x_{\text{c}}$ of the equation
\cite{Rezzolla_Zhidenko14},
\begin{equation} \label{eq:x_Photon_Sphere_Equation}
(1-x) - \frac{N(x)}{\partial_x N(x)} = 0\,,
\end{equation}
where $\partial_x$ denotes a derivative w.r.t. $x$. We denote by
$x_{\text{ps}}$ the location of the absolute maximum of the null geodesic
potential, and the corresponding impact factor of this photon
$\xi_{\text{ps}}$ is given as,
\begin{equation}
\xi_{\text{ps}} = \frac{r_0}{(1-x_{\text{ps}})N(x_{\text{ps}})}\,.
\end{equation}  

\subsubsection{Orbital angular velocity on stable circular geodesics}

The class of equatorial Kepler observers in a static spacetime
\eqref{eq:General_Static_Metric} correspond to stable circular timelike
geodesic motion, and satisfy, as discussed above,
\begin{equation}
0 = \frac{\tilde{\xi}^2}{r^2} - \frac{1}{N^2} + \frac{1}{\tilde{E}^2}\,, \ \
0 = \frac{\tilde{\xi}^2}{r^3} - \frac{\partial_r N}{N^3}\,.
\end{equation}
From the above, we can straightforwardly obtain the associated equatorial
Kepler frequency at a given radius $\Omega_{\text{K}} =
\dot{\phi}/\dot{t}$ as,
\begin{equation} \label{eq:Kepler_Frequency}
\Omega_{\text{K}} := \frac{\tilde{\xi}N^2}{r^2} = \sqrt{\frac{N(r)\partial_r N(r)}{r}}\,.
\end{equation}
Furthermore, since around sufficiently massive black holes, pulsars
(rotating neutron stars that spin around their axes, and emit radiation)
can be treated as test objects and are visible to fixed asymptotic
observers, measuring the rate at which pulses from them are recorded on
earth can be useful towards setting up strong field tests of GR, since
this rate depends on the properties of its motion. From pulse profiles of
pulsars moving in the vicinity of static black holes, the orbital angular
frequency can potentially be extracted, and since this frequency depends
on the properties of the central compact object like the mass and charge
of the central object, one could in principle extract these parameters
for the spacetime as well \cite{Kocherlakota+19}.

A particle moving on the ISCO corresponds to the absolute minima of
$\tilde{V}_{\text{eff}}$, and satisfies additionally
$\partial^2_r\tilde{V}_{\text{eff}} = 0$ i.e.,
\begin{equation}
\frac{\tilde{\xi}^2}{r^4} - \frac{(\partial_r N)^2}{N^4} + \frac{\partial_r^2 N}{3 N^3} = 0\,.
\end{equation}
Clearly, the ISCO is also only marginally stable. The ISCO radius then is
the solution of \cite{Rezzolla_Zhidenko14, DeLaurentis+18},
\begin{equation} \label{eq:ISCO_Equation}
3 N \partial_r N - 3 r (\partial_r N)^2 + r N \partial_r^2 N= 0\,,
\end{equation}
and the corresponding orbital angular velocity is given as
$\Omega_{_{\text{ISCO}}} = \Omega_{\text{K}}(r_{_{\text{ISCO}}})$. Kepler
observers exist only outside the ISCO i.e., only for $r \geq
r_{_{\text{ISCO}}}$.

In the current parametrization scheme, the Kepler orbital angular
velocity is given by,
\begin{equation} \label{eq:Opr_x}
\Omega_{\text{K}} = \frac{\sqrt{(1-x)^3 N(x) \partial_x N(x)}}{r_0}\,,
\end{equation}
and the ISCO location is given as $r_{_{\text{ISCO}}} = r_0/(1-x_{_{\text{ISCO}}})$, where $x_{_{\text{ISCO}}}$ satisfies,
\begin{equation} \label{eq:x_ISCO_Equation}
N \partial_x N - 3(1-x)(\partial_x N)^2 + (1-x) N \partial_x^2 N = 0\,.
\end{equation}
Finally, the ISCO angular velocity is obtained from equation
\eqref{eq:Opr_x} by setting $x = x_{_{\text{ISCO}}}$.

\subsubsection{Strong gravitational lensing}
We study now the lensing properties of compact objects within this
parametrization framework. For equatorial null geodesics, we can
characterize the deflection due to gravity via
\begin{equation}
\frac{\text{d}\phi}{\text{d}r} = \frac{\dot{\phi}}{\dot{r}} =
\pm\frac{{L}/{r^2}}{({N}/{B})\sqrt{{E^2}/{N^2} -
    {L^2}/{r^2}}} = \pm\frac{1}{r^2}\frac{B}{\sqrt{1/\xi^2 -
    N^2/r^2}}\,,
\end{equation}
where the sign $+$ or $-$ is determined by whether it is out-going
($\dot{r} \! > \! 0$) or in-going ($\dot{r} \! < \! 0$) respectively. For
a null geodesic starting from and ending at asymptotic infinity, the
point where it is closest to the compact object, namely its turning point
$r  =  r_{\text{tp}}$, is obtained from the condition that $\dot{r} =
0$ there, which gives,
\begin{equation}
\xi = \frac{r_{\text{tp}}}{N(r_{\text{tp}})}\,.
\end{equation}
Then, the total deflection due to gravitational lensing
$\Delta\phi_{\text{GL}}(r_{\text{tp}})$ of such a null geodesic, i.e.,
its deviation from a straight line, is given as \cite{Weinberg72},
\begin{equation}
\Delta\phi_{\text{GL}}(r_{\text{tp}}) =
2\left|\int^{\infty}_{r_{\text{tp}}}\frac{\text{d}r}{r^2}\frac{B(r)}{\sqrt{{N^2(r_{\text{tp}})}/{r_{\text{tp}}^2}
    - {N^2(r)}/{r^2}}}\right| - \pi\,.
\end{equation}
Within the current parametrization scheme, this may be rewritten as,
\begin{equation} \label{eq:Lensing_x}
\Delta\phi_{\text{GL}}(x_{\text{tp}}) = 2\left|\int^{1}_{x_{\text{tp}}}\text{d}x\frac{B(x)}{\sqrt{(1-x_{\text{tp}})^2N^2(x_{\text{tp}}) - (1- x)^2N^2(x)}}\right| - \pi\,,
\end{equation}
where $x_{\text{tp}} := 1 - {r_0}/{r_{\text{tp}}}$. This integral is
finite only if the turning point lies outside the photon sphere, i.e.,
$x_{\text{ps}} \! < \! x_{\text{tp}} \! < \! 1$.

In the next section, we display the parameters necessary to parametrize
various spacetimes based on the parametrization scheme described in
Sec. \ref{sec:eRZ_Parametrization} and then proceed to demonstrate how
efficiently metric functions and the various observables discussed in
this section are characterized in this framework.

\section{Characterizing spacetimes and observables in the parametrization scheme}
\label{sec:Spacetimes_eRZ}
We now discuss the conventions used here and the layout of this section
before we enter into a brief description of the various BH, boson star,
and naked singularity spacetimes considered in this work.

We employ geometrized units throughout $8\pi G = c = 1$. Deviations in
the gravitational constant $G$ or the Planck length $l_{\text{p}}$ can be
measured in scales of their canonical values. Further, since the
spacetimes considered here are all asymptotically-flat, the ADM mass $M$
can be used to fix a length scale for the Schwarzschild-like coordinate
system used in equation \eqref{eq:General_Static_Metric_RZ}. If we switch
to a mass-dimensionless radial coordinate $\bar{r} = r/M$, a direct
comparison of various quantities (observables and metric functions)
associated with various solutions becomes meaningful.

This also allows us to obtain the dependence of various observables
associated with the solutions considered here on the other relevant
physical ``charges'', like the scalar or electric or magnetic charge
etc. In instances when the mass of a compact object has been ascertained
from observations to requisite precision, one could then potentially look
for the dependence on other ADM charges of observational data. The
mass-scaling of the various observables considered here is clear from
Sec. \ref{sec:Observables_eRZ}. The impact parameter of a photon on an
unstable circular orbit $\xi_{\text{ps}}$, the orbital angular frequency
for Kepler observers $\Omega_{\text{K}}$, and the deflection angle due to
gravitational lensing $\Delta\phi_{\text{GL}}$ scale with mass as $M^2,
M, M^{-1}$ and $M^0$ respectively. Further since the conformal parameter
$x$ is scale-invariant, the metric functions $N^2(x)$ and $B^2(x)$ are
unaffected i.e., changing the units of the radial coordinate does not
affect the Pad\'e expansion coefficients.  This is an important quality
that makes the definition of a parametrization space $\Pi$ as in
Sec. \ref{sec:Parametrization_Space} useful.

For easy access, the BH metric functions used here have been compiled in
table \ref{table:BH_Spacetimes}. We display in table
\ref{table:eRZ_Parameters} the parametrization coefficients up to fourth
order ($\epsilon, a_i, b_i$ for $0 \! \leq \! i \! \leq \! 4$) for all
solutions considered here, BHs and otherwise. As discussed above, $n_0 \!
= \! 0$ for BH spacetimes and $n_0 = \epsilon$ for non-BH spacetimes
since we set the inner boundary in these cases to correspond to the
Schwarzschild radius, $\bar{r}_0 = 2$.

In the columns under part I of table \ref{table:Numerical_Results}, we
show the relative error in obtaining $\xi_{\text{ps}}$ and
$\Omega_{_{\text{ISCO}}}$ for various spacetimes, when using Pad\'e
approximants truncated at the fourth order ($a_5 = b_5 = 0$), as an
indicative quantitative measure of the `goodness' of the current
parametrization scheme. For instance, for a Bardeen BH with specific
magnetic charge $\bar{q}_{\text{m}} = 0.75$, we get $|1 - \xi_{\text{ph};
  a_5 = 0}/\xi_{\text{ph; exact}}|$ to be $7.35 \times 10^{-6}$. We also
show the maximum relative error in obtaining the metric functions
$N^2(x)$ and $B^2(x)$ over the entire range $0 \! \leq x \! < \!
1$. Finally, we display also the maximum relative error in approximating
the orbital angular frequency of Kepler observers $\Omega_{\text{K}}$ and
the deflection angle due to gravitational lensing
$\Delta\phi_{\text{GL}}$ over the entire accretion disk
$x_{_{\text{ISCO}}} \! < \! x \! < \! 1$, where $x_{_{\text{ISCO}}}$ is
defined via equation \eqref{eq:x_ISCO_Equation}.

To compare the goodness of the present approximation, we show the
\textit{exact} relative differences from the Schwarzschild values of
$\xi_{\text{ps}}$ and $\Omega_{_{\text{ISCO}}}$ for various spacetimes
under part II of table \ref{table:Numerical_Results}. For example, under
the column for impact parameters, we report $|1 - \xi_{\text{ps;
    Spacetime}}/\xi_{\text{ps; Schwarzschild}}|$. We also show the
relative error in obtaining the deflection angle due to gravitational
lensing at the ISCO radius $\Delta\phi_{\text{GL}}(r_{_{\text{ISCO}}})$
there.

Since the relative error levels in obtaining the exact observables within
this parametrization scheme are significantly lower when compared to the
deviation of their exact values from the Schwarzschild spacetime, setting
up precision tests is possible. Furthermore, since the number of
parameters to characterize the wide variety of compact objects in use
here is small (11), we conclude that this parametrization scheme is a
promising framework to test theories of gravity and the quantum field
theoretic effects that may show up in astrophysical data related to
compact objects. It is remarkable that this parametrization method
performs quite well across the entire radial patch, and allows one to
capture both weak- and strong-gravitational field regimes simultaneously.

\subsection{Parametrization Space $\Pi$}
\label{sec:Parametrization_Space}

We now introduce the geometric notion of a parametrization space. If we
think of each set of PPN and Pad\'e expansion coefficients $(\epsilon,
a_{0 \leq i \leq 4}, b_{0 \leq i \leq 4})$ as being points of some
abstract `parametrization space' $\Pi$, then it is clear that for each
set of physical charges $q_j$ for a given spacetime, we can associate a
point $\pi(q_j) \in \Pi$. As we vary the physical parameters $q_j$
associated with that particular spacetime over its entire range, we
obtain a curve or surface in $\Pi$, depending on the number of charges,
$q_1, q_2 \cdots q_j$. We can then use the usual Euclidean
$\mathscr{L}^2$-norm on $\Pi$ to measure distances between such curves or
surfaces, or equivalently between solutions. In particular, we define the
deviation of a solution from the Schwarzschild BH spacetime, which sits
at the origin of this space, as simply being given by,
\begin{equation}
\mathscr{L}^2_0 := (\pi-0)^2 = \left(\epsilon^2 + \sum_{i=0}^{4}(a_i^2 + b_i^2)\right)^{1/2}\,.
\end{equation}
Table \ref{table:eRZ_Parameters} then lists the coordinates of various
spacetimes in this space. When two `solution curves' intersect, the
corresponding spacetimes match approximately at the common point. It is
to be noted that since we have used only the first few ($n \leq 4$)
expansion coefficients to set up $\Pi$, various spacetimes are
approximated at varying degrees of accuracy. However since a higher-order
approximation does not affect low-order PPN or Pad\'e coefficients, we
can always compare spacetimes meaningfully on $\Pi$. For an alternative
prescription to measure differences between solutions one may see
\cite{Suvorov20}, where a superspace approach was adopted.

Now, for each solution, we use a grid with 1000 points for each physical
charge, and obtain the PPN and Pad\'e coefficients $\pi(q)$ at each grid
point $q$. For the Bardeen BH, there is a single physical parameter
$\bar{q}_{\text{m}}$ and we obtain obtain $\pi(\bar{q}_{\text{m}})$ for
1000 points between $0 \leq \bar{q}_{\text{m}} \lesssim .76$. We then
ascertain whether the PPN constraints \eqref{eq:PPN_RZ_Param} are met at
each grid point and thus obtain the PPN-allowed parameter values for each
spacetime.  This is reported in table \ref{table:BH_Spacetimes}. We
introduce another useful quantity, the $\mathscr{L}^\infty$-norm, which
is defined as,
\begin{align}
\mathscr{L}^\infty :=&\ \text{max}\left\{|\epsilon|, |a_i|, |b_i|\ (0 \leq i \leq 4)\right\}\,,
\end{align}
to characterize deviations from the Schwarzschild BH solution, and obtain
its value at each grid point for a particular spacetime. We can then find
the maximum value of $\mathscr{L}^\infty_{\text{max}}$ over all the
(charge) grid points for a particular spacetime, to obtain a measure of
the extent of the region in $\Pi$ that the spacetime has support
on. Since all of these solutions become approximately Schwarzschild in
the limit of approach to a particular physical parameter value, as can be
seen from table \ref{table:BH_Spacetimes},
$\mathscr{L}^\infty_{\text{max}}$ gives a sense of the maximal deviation
from the Schwarzschild BH solution. Essentially, if one samples this
range of the parameter space along all axes, one is sure to have
characterised that particular spacetime. We report
$\mathscr{L}^\infty_{\text{max}}$ for each spacetime in the last column
on table \ref{table:BH_Spacetimes} for BHs. Note that for all of the
spacetimes considered here, this quantity
$\mathscr{L}^\infty_{\text{max}}$ is finite. That is, by sampling the
region $0 < |\epsilon|, |a_i|, |b_i|\ (0 \leq i \leq 4) < 10$, we have
completely characterised all of the BH spacetimes used in this work. Of
course, since the exact value of $\mathscr{L}^\infty_{\text{max}}$
depends on the resolution of the grid, we report here a rounded up value
as an indicative measure.

For the JNW naked singularity spacetime, we obtain
$\mathscr{L}^\infty_{\text{max}} \approx 26$ (we use a very coarse grid
$\nu = 0.1, 0.2, \cdots , 0.9$ for this spacetime). For each of the boson
star models considered here, this number appears to be around
$\mathscr{L}^\infty_{\text{max}} \approx 10^3$. What this means is that
all of the spherically symmetric metrics used here lie in a compact
region of $\Pi$ around the origin.

Note that we will not restrict our study to the PPN-allowed ranges of the
physical charges for the BH spacetimes, reported in table
\ref{table:BH_Spacetimes}, but explore their entire ranges instead.

\subsection{Black Holes}

\begin{center}
\begin{table*} 
\caption{Metric functions, $N^2(\bar{r})$ and $B^2(\bar{r})$, of the BH
  spacetimes from arbitrary theories of gravity that we have
  considered. Here, $\bar{r} = r/M$ and $M$ is the ADM mass. The location
  of the Killing horizon $\bar{r}_0$ is obtained by solving
  $N^2(\bar{r}_0) = 0$, and is used in defining the conformal radial
  coordinate $x = 1 - \bar{r}_0/\bar{r}$. The patch of the spacetime that
  we capture within this parametrization scheme is the entire exterior
  horizon geometry, $0 \leq x < 1$. It is to be noted that for the
  Einstein-aether BHs, the Killing horizon is different from the actual
  causal boundary of the BH region. Also, the term $\bar{r}_-$ appearing
  in $B^2$ for the Modified Hayward BH is the smaller zero of its $N^2$
  metric function. We also show below the PPN-allowed range of the
  relevant parameter for each spacetime. Finally, in the last column we
  display the (rounded-up) maximum of the absolute values of all
  expansion coefficients for all PPN-allowed parameter values for a given
  spacetime, $\mathscr{L}^\infty_{\text{max}}$, with its order of
  magnitude given in square brackets.  This number is meant to provide a
  rough sense of the size of the region of this 11D parameter space on
  which a particular spacetime of astrophysical interest has support.}
\label{table:BH_Spacetimes}
\renewcommand{\arraystretch}{2.5}
\begin{tabular}[t]{|c|c|c|c|c|c|}
\hline
{Spacetime} & {Physical Charge} & ${N^2 = -g_{00}}$ & ${B^2 = -g_{00}g_{11}}$ & {PPN Constrained} & $\mathscr{L}^\infty_{\text{max}}$ \\
\hline
RN \cite{Reissner16_Nordstrom18} & $0 < \bar{q} \leq 1$ & $1 - \frac{2}{\bar{r}} + \frac{\bar{q}^2}{\bar{r}^2}$ & 1 & $0 < \bar{q} \lesssim 2.1 \times 10^{-2}$ & 1 [-4]  \\
\hline
E-ae 2 \cite{Berglund+12}  & $0 < c_{13} < 1,$ & $1 - \frac{2-c_{14}}{\bar{r}} - \frac{(2 c_{13} - c_{14})(2-c_{14})^2}{8(1-c_{13})}\frac{1}{\bar{r}^2} $ & 1 & $0 < c_{13} < 1,$ & 8 [-1] \\
 & $0 \leq c_{14} \leq 2 c_{13} < 2$ & & & $0 \leq c_{14} \leq 2 c_{13} < 2$ & \\
\hline
E-ae 1 \cite{Berglund+12} & $0 < c_{13} < 1$ & $1 - \frac{2}{\bar{r}} - \frac{3^3c_{13}}{2^4(1 - c_{13})}\frac{1}{\bar{r}^4}$ & 1 & $0 < c_{13} < 1$ & 6 [-1] \\
\hline
Bardeen \cite{Bardeen68} & $0 < \bar{q}_{\text{m}} \leq \sqrt{16/27}$ & $1 - \frac{2\bar{r}^2}{\left(\bar{r}^2 + \bar{q}_{\text{m}}^2\right)^{3/2}}$ & 1 & $0 < \bar{q}_{\text{m}} \leq \sqrt{16/27}$ & 1 [1] \\
\hline
Hayward \cite{Hayward06, Held+19} & $0 < \bar{l} \leq \sqrt{16/27}$ & $1 - \frac{2\bar{r}^2}{\bar{r}^3 + 2 \bar{l}^2}$ & 1 & $0 < \bar{l} \leq \sqrt{16/27}$  & 6 [0] \\
\hline
Bronnikov \cite{Bronnikov01} & $0 < \bar{q}_{\text{m}} \lesssim 1.05$ & $1- \frac{2}{\bar{r}}\left(1 - \tanh{\frac{\bar{q}_{\text{m}}^2}{2\bar{r}}}\right)$ & 1 & $0 < \bar{q}_{\text{m}} \lesssim 1.05$ & 2 [0] \\
\hline
EEH \cite{Yajima_Tamaki01} & $0 < \bar{\alpha},$ & $1 - \frac{2}{\bar{r}} + \frac{\bar{q}_{\text{m}}^2}{\bar{r}^2} - \bar{\alpha}\frac{2\bar{q}_{\text{m}}^4}{5\bar{r}^6}$ & 1 &  $0 < \bar{\alpha},$  & 1 [0] \\
 & $0 < \bar{q}_{\text{m}}$ & & & $0 < \bar{q}_{\text{m}} \lesssim 2.1 \times 10^{-2}$ & \\
\hline
Frolov \cite{Frolov16} & $0 < \bar{l} \leq \sqrt{16/27}$, & $1 - \frac{(2\bar{r} - \bar{q}^2)\bar{r}^2}{\bar{r}^4 + (2\bar{r} + \bar{q}^2)\bar{l}^2}$ & 1 & $0 < \bar{l} \leq \sqrt{16/27},$  & 4 [0] \\
& $0 \leq \bar{q} \leq 1$ & & & $0 < \bar{q} \lesssim 2.1 \times 10^{-2}$ & \\
\hline
KS \cite{Kazakov_Solodhukin94} & $0 < \bar{a}$ & $-\frac{2}{\bar{r}} + \frac{\sqrt{\bar{r}^2 - \bar{a}^2}}{\bar{r}}$ & 1 &  $0 < \bar{a} \lesssim 3.0 \times 10^{-2}$  & 4 [-1] \\
\hline
CFM A \cite{Casadio+02} & $\beta < 1$ & $1 - \frac{2}{\bar{r}}$ & $\left(1 - \frac{3}{2\bar{r}}\right)\left(1 - \frac{4\beta - 1}{2\bar{r}}\right)^{-1}$ & $|\beta - 1| \lesssim 2.3 \times 10^{-5}$  & 3 [0] \\
\hline
CFM B \cite{Casadio+02} & $1 < \beta < 5/4$ & $1 - \frac{2}{\bar{r}}$ & $\left(1 - \frac{3}{2\bar{r}}\right)\left(1 - \frac{4\beta - 1}{2\bar{r}}\right)^{-1}$ & $|\beta - 1| \lesssim 2.3 \times 10^{-5}$  & 3 [0] \\
\hline
Mod. Hayward \cite{Frolov16} & $0 < \bar{l} \leq \sqrt{16/27}$ & $1 - \frac{2\bar{r}^2}{\bar{r}^3 + 2 \bar{l}^2}$ & $\frac{\bar{r}^6 + \bar{r}_-^6}{\bar{r}^6 + \bar{r}_{\text{H}}^4 \bar{r}_-^2}$ &  $0 < \bar{l} \leq \sqrt{16/27}$  & 6 [0] \\
\hline
EMd \cite{Gibbons_Maeda88, Garfinkle+92, Garcia+95} & $0 < \bar{q} \leq \sqrt{2}$ & $1 - \frac{\sqrt{4\bar{r}^2 + \bar{q}^4} - \bar{q}^2}{\bar{r}^2}$ & $\frac{4\bar{r}^2}{4\bar{r}^2 + \bar{q}^4}$ & $0 < \bar{q} \lesssim 2.1 \times 10^{-2}$  & 2 [0] \\
\hline
\end{tabular}
\end{table*}
\end{center}

Note that since the mass $M$ is the ADM mass for all of the solutions
below, and is a free parameter; We will only discuss the remaining
charges in what follows.

The Reissner-Nordstr\"om (RN; \cite{Reissner16_Nordstrom18}) BH describes
a charged BH in GR, with specific charge $0 < \bar{q} \leq 1$.

We consider two BH solutions reported in \cite{Berglund+12} that are
obtained from the Einstein-aether (E-ae) Lagrangian. In an aether theory,
LLI is violated due to the existence of the aether vector field. The
first of the two solutions, which we call the E-ae 1 BH is a single
parameter solution, $0 \! < \! c_{13} \! < \! 1$, and the E-ae 2 solution
represents a two-parameter family of BHs which take values, $0 \! < \!
c_{13} \! < \! 1$ and $0 \! \leq c_{14} \! \leq 2c_{13} \! < \! 2$. Here
$c_{13}$ and $c_{14}$ are coupling constants that control the aether
Lagrangian. It is to be noted that for these spacetimes the causal
horizons that separate the BH interiors $B \equiv M - J^-(I^+)$ from
their exteriors, called the universal horizons in \cite{Berglund+12}, are
different from Killing horizons and we use the latter when defining the
conformal coordinate $x$.

The Einstein-gravity Lagrangian when coupled to a particular non-linear
electrodynamics (NLED) Lagrangian $\mathcal{L}(F)$, which reduces to
Maxwell in the weak-field limit, with $F$ the electromagnetic
field-strength scalar (see equation 29 of \cite{Bronnikov01}; see also
\cite{Ayon-Beato_Garcia99}), yields regular, magnetically-charged BH
solutions. These Bronnikov BH solutions are given by equations (3, 11,
30) of \cite{Bronnikov01}, with the specific magnetic charge, $0 \! < \!
\bar{q}_{\text{m}} \! \lesssim \! 1.05$, as the only additional charge.

When considering the Einstein Lagrangian coupled to the Euler-Heisenberg
(EH) NLED Lagrangian, which is considered to be an effective action of a
superstring theory \cite{Bern_Morgan94}, one can obtain a
magnetically-charged BH solution \cite{Yajima_Tamaki01}. This
Einstein-Euler-Heisenberg (EEH) BH depends on two parameters, $0 \! <
\bar{\alpha}$ and $0 \! < \! \bar{q}_{\text{m}}$, the former of which is
the coupling constant of the $F^2$ piece of the EH Lagrangian and is
expected to be determined by the string tension $\alpha^\prime$
\cite{Yajima_Tamaki01}.

It is important to note that due to the self-interaction introduced by
the non-linearity of the NLED Lagrangians in the Bronnikov and the EEH BH
solutions, photons do not propagate along null geodesics of
\eqref{eq:General_Static_Metric_RZ} (see, e.g., the discussion in
\cite{Stehle_DeBaryshe66}). However, as was discussed in
\cite{Cuzinatto+15}, the event horizons are still determined by the
zeroes of the null expansions of
\eqref{eq:General_Static_Metric_RZ}. Therefore, our definition of the
conformal coordinate $x$ is unchanged, and we can still characterize
these BH spacetimes within the current parametrization scheme. However,
other important phenomena such as geometric redshift and light deflection
are modified by the NLED Lagrangian. While we are able to show that these
solutions are obtained within our parametrization scheme to very high
accuracy, and also show that the errors in obtaining the ISCO frequency
and the Kepler frequency are also very small (matter is still minimally
coupled), we find studying the accuracy in obtaining the photon sphere
impact parameter or the deflection of photons due to gravitational
lensing for these spacetimes to be beyond the scope of the current
article\footnote{In the case of the Bronnikov BH, while it has been
  discussed that NLED photons propagate along the null geodesics of an
  effective metric given in equations (26, 27) of \cite{Bronnikov01} (see
  also \cite{Novello+00a, Novello+00b}), this supplementary `optical
  metric' is plagued by coordinate and curvature singularities, and the
  causal structure of this spacetime is unclear to us.}.
 
The Bardeen-BH model, proposed in \cite{Bardeen68}, is the result of the
collapse of charged matter, with the usual central singularity replaced
by a regular charged matter core. The only relevant parameter in this
solution takes values $0 \! < \! \bar{q}_{\text{m}} \! \lesssim \! 0.77$.
More recently, it was shown in \cite{Ayon-Beato_Garcia00} that this BH
can also be obtained as an exact magnetically-charged solution of an
Einstein-NLED Lagrangian.

The Hayward BH model \cite{Hayward06} proposes a method to resolve the
central singularity in uncharged BHs in GR by adding a region with
positive cosmological constant $\Lambda = 3/l^2$ (de Sitter) close to the
centre, where $l$ is the Hubble length. Such a model is expected to be
justified by the properties of matter \cite{Sakharov66, Gliner66} or the
quantum theory of gravity \cite{Markov82, Frolov+90,
  Mukhanov_Brandenberger92, Brandenberger+93} close to the centre of the
BH. While $l$ provides a length scale for when such effects might set in,
and can therefore be related to the Planck length, larger length scales
are not strictly excluded. We will consider here the entire range of the
parameter for which BH solutions are admitted, $0 \! < \! \bar{l} \!
\lesssim \! 0.77$. Since we have introduced the ADM mass into the
definition of $\bar{l}$, which can determined in terms of the canonical
values of $G$ and $c$, fixing a particular length scale $l$ can be
thought of equivalently as considering BHs within a certain mass range.

The charged generalisation of the Hayward model given by equation (4.1)
of \cite{Frolov16} is referred to as the Frolov BH here and has an
additional parameter which takes values $0 \! < \! \bar{q} \! \leq \!
1$. Another generalisation of the Hayward model is also presented there,
which modifies the redshift function in equations (2.47, 2.50); This we
refer to as the modified Hayward model.

The effective dynamics of spherically symmetric fluctuations of the 4D
gravitational field can be shown to be governed by a 2D dilaton gravity
action \cite{Kazakov_Solodhukin94}. By integrating out these
fluctuations, one can obtain the (approximate) Kazakov-Solodhukin (KS) BH
metric, given in equation (3.18) of \cite{Kazakov_Solodhukin94}. The
relevant parameter for this solution takes values $0 < a$, and determines
the area of the singular two-sphere, i.e., $\mathcal{A}_{\text{sing}} =
4\pi a^2$. While this parameter should be roughly of the order of the
Planck length, we allow it to take all positive values here. While, more
significantly, non-singular solutions are also presented there, we do not
consider them here.

Projecting the 5D vacuum Einstein equations onto a time-like manifold of
codimension one (brane) yields the usual ADM Hamiltonian and momentum
constraints, for spherically symmetric solutions. If one chooses the
four-metric on the brane to be given by equation
\eqref{eq:General_Static_Metric_RZ} with $N^2(\bar{r}) = (1-2/\bar{r})$,
this Hamiltonian constraint equation uniquely determines the other metric
function, and a one-parameter family of Casadio-Fabbri-Mazzacurati (CFM)
BH solutions are obtained, given in equation (8) of
\cite{Casadio+02}. For $\beta < 1$ these are the singular (CFM A) and for
$1 < \beta < 5/4$ these are non-singular (CFM B). The CFM B BHs in fact
contain traversable wormholes (the minimal sphere is behind the horizon;
see figure 2 of \cite{Casadio+02}). The parameter $\beta$ here
corresponds exactly to the PPN $\beta$ parameter.

Due to the coupling of the dilaton field to the electromagnetic field
strength $\boldsymbol{F}$ in heterotic string theory, the
Einstein-Maxwell-dilaton (EMd) BH is the appropriate electromagnetically
charged BH solution in the low-energy limit for this theory
\cite{Gibbons_Maeda88, Garfinkle+92, Garcia+95}, as opposed to the RN
BH. The EMd BH is characterised by the boundary value of the dilaton
field $\phi_0$ and the specific electric or magnetic charge
$\bar{q}$. For convenience, we consider here solutions with $\phi_0 =
0$. In this case, EMd BH solutions exist for $0 < \bar{q} \leq \sqrt{2}$.

All BH solutions (barring the NLED BHs, which we have not studied here)
cast shadows and all BH solutions admit ISCOs.

\begin{center}
\begin{table*}
\caption{PPN and Pad\'e approximant coefficients up to order four for
  various spacetimes. As discussed in the following table
  \ref{table:Numerical_Results}, at this order, the relative errors in
  obtaining both the metric functions and various observables is at the
  level of about $10^{-6}$, and in fact systematically much lower for
  BHs. For the boson star and naked singularity, we have used $\bar{r}_0 =
  2$, and so $n_0 = \epsilon$. On the other hand, for BHs, $n_0 = 0$. The
  parameter $\epsilon = 2/\bar{r}_0 - 1$ measures the
  difference of the horizon radius from its Schwarzschild radius for
  BHs. Also, as was discussed in Sec. \ref{sec:PPN_Constraints} above,
  $\epsilon, a_0, b_0$ are the only relevant parameters to test whether a
  particular spacetime is ``PPN allowed.'' Finally, if a Pad\'e
  coefficient of some order $n$ vanishes, i.e., if $a_{n \geq 1} = 0$ or
  $b_{n \geq 1} = 0$, then the corresponding metric function, $N^2$ or $B^2$, is exactly characterised
  within this approximation scheme at order-$n$ (see, e.g., the case of E-ae 
  1 BHs below).}
\label{table:eRZ_Parameters}
\renewcommand{\arraystretch}{1.3}
\begin{tabular}[t]{|c|c| |c|c|c| |c|c|c|c| |c|c|c|c||}
\hline
Black Hole & Physical & \multicolumn{3}{c||}{PPN coefficients (Taylor)} & \multicolumn{4}{c||}{Higher Order $a$-coefficients (Pad\'e)} & \multicolumn{4}{c||}{Higher Order $b$-coefficients (Pad\'e)} \\
\cline{3-13}
Spacetime & Charge & $\epsilon$ & $a_0$ & $b_0$ & $a_1$ & $a_2$ & $a_3$ & $a_4$ & $b_1$ & $b_2$ & $b_3$ & $b_4$ \\
\hline
Schwarzschild & - & 0 & 0 & 0 & 0 & 0 & 0 & 0 & 0 & 0 & 0 & 0 \\
\hline
RN & $\bar{q} = 0.5$ & 0.07180 & 0.07180 & 0 & 0 & 0 & 0 & 0 & 0 & 0 & 0 & 0 \\
\cline{2-13}
& $\bar{q} = 0.9$ & 0.39286 & 0.39286 & 0 & 0 & 0 & 0 & 0 & 0 & 0 & 0 & 0 \\
\hline
E-ae 2 & [0.1, 0.1] & -0.01352 & -0.01352 & 0 & 0 & 0 & 0 & 0 & 0 & 0 & 0 & 0 \\
\cline{2-13}
[$c_{13}, c_{14}$]  & [0.9, 0.1] & -0.51007 & -0.51007 & 0 & 0 & 0 & 0 & 0 & 0 & 0 & 0 & 0 \\
\cline{2-13}
& [0.9, 1.7] & -0.10102 & -0.10102 & 0 & 0 & 0 & 0 & 0 & 0 & 0 & 0 & 0 \\
\hline
E-ae 1 & $c_{13} = 0.5$ & -0.07666 & 0 & 0 & 0.07666 & 0 & 0 & 0 & 0 & 0 & 0 & 0 \\
\cline{2-13}
& $c_{13} = 0.9$ & -0.26982 & 0 & 0 & 0.26982 & 0 & 0 & 0 & 0 & 0 & 0 & 0 \\
\hline
Bardeen & $\bar{q}_{\text{m}} = 0.25$ & 0.02471 & 0 & 0 & 0.00100 & 0.46236 & -0.54081 & 0.04101 & 0 & 0 & 0 & 0 \\
\cline{2-13}
& $\bar{q}_{\text{m}} = 0.75$ & 0.53960 & 0 & 0 & 0.32920 & -0.07610 & 3.42051 & -3.89389 & 0 & 0 & 0 & 0 \\
\hline
Hayward & $\bar{l} = 0.25$ & 0.01641 & 0 & 0 & -0.01561 & -0.09932 & 0.70929 & -0.40533 & 0 & 0 & 0 & 0 \\
\cline{2-13}
& $\bar{l} = 0.75$ & 0.33333 & 0 & 0 & -0.08333 & -3.75000 & 3.46667 & -0.15897 & 0 & 0 & 0 & 0 \\
\hline
Bronnikov & $\bar{q}_{\text{m}} = 0.5$ & 0.07167 & 0.07178 & 0 & 0.00011 & -0.00538 & 0.33468 & -0.33299 & 0 & 0 & 0 & 0 \\
\cline{2-13}
& $\bar{q}_{\text{m}} = 1.05$  & 1.03616 & 1.14273 & 0 & 0.08279 & -0.36596 & 0.44373 & -0.30767 & 0 & 0 & 0 & 0 \\
\hline
EEH & [1, 0.05] & 0.79539 & 0.80585 & 0 & 0.03140 & 1.00000 & -0.66667 & 0.16667 & 0 & 0 & 0 & 0 \\
\cline{2-13}
$[\bar{q}_{\text{m}}, \bar{\alpha}]$ & [1,1] & 0.48364 & 0.55030 & 0 & 0.19997 & 1.00000 & -0.66667 & 0.16667 & 0 & 0 & 0 & 0 \\
\hline
Frolov & [0.5, 0.25] & 0.09732 & 0.07526 & 0 & -0.02039 & -0.15602 & 0.74279 & -0.37350 & 0 & 0 & 0 & 0 \\
\cline{2-13}
[$\bar{q}, \bar{l}$]  & [0.5, 0.6] & 0.36454 & 0.11637 & 0 & -0.09263 & -2.39466 & 2.33431 & -0.20990 & 0 & 0 & 0 & 0 \\
\cline{2-13}
& [0.9, 0.25] & 0.61799 & 0.53013 & 0 & -0.04369 & -1.59541 & 1.85133 & -0.17365 & 0 & 0 & 0 & 0 \\
\hline
KS & $\bar{a} = 1$ & -0.10557 & -0.10000 & 0 & 0.00689 & 0.34416 & 0.32590 & -0.22973 & 0 & 0 & 0 & 0 \\
\cline{2-13}
& $\bar{a} = 10$ & -0.80388 & -0.48077 & 0 & 2.97202 & 17.9580 & 8.72455 & 16.6646 & 0 & 0 & 0 & 0 \\
\hline
CFM A & $\beta = -0.9$ & 0 & 0 & -0.95000 & 0 & 0 & 0 & 0 & 0.29100 & -0.80649 & 0.94227 & 1.09107 \\
\cline{2-13}
& $\beta = 0.9$ & 0 & 0 & -0.05000 & 0 & 0 & 0 & 0 & -0.10485 & 2.12935 & 0.03647 & 2.39174 \\
\hline
CFM B & $\beta = 1.1$ & 0 & 0 & 0.05000 & 0 & 0 & 0 & 0 & 0.24099 & 4.93512 & 0.09874 & 4.23521 \\
\cline{2-13}
& $\beta = 1.2$ & 0 & 0 & 0.10000 & 0 & 0 & 0 & 0 & 1.13607 & 13.6579 & 1.56198 & 9.44727 \\
\hline
Mod. Hayward & $\bar{l} = 0.25$ & 0.01641 & 0 & 0 & -0.01561 & -0.09932 & 0.70929 & -0.40533 & -0.00919 & 3.91749 & -2.52087 & 0.52343 \\
\cline{2-13}
& $\bar{l} = 0.75$ & 0.33333 & 0 & 0 & -0.08333 & -3.75000 & 3.46667 & -0.15897 & -0.12939 & 2.05606 & -3.40434 & 1.48027 \\
\hline
EMd & $\bar{q} = 0.7$ & 0.15087 & 0.16225 & 0 & -0.00011 & 0.48046 & -0.52013 & 0.01976 & -0.00979 & -0.02928 & 0.49677 & -0.50314 \\
\cline{2-13}
& $\bar{q} = 1.4$ & 6.07107 & 24.5000 & 0 & -13.3186 & -0.68426 & 0.05403 & -0.99713 & -0.72270 & -1.64581 & 0.44380 & -0.45354 \\
\hline
MBS A & - & 0.25607 & -2860.02 & 0 & 2860.11 & -0.99991 & -0.00005 & -1.03642 & -0.33233 & -0.54166 & 1.75930 & -2.45991 \\
\hline
MBS B & - & 0.49436 & -2860.02 & 0 & 2860.58 & -0.99991 & 0.00007 & -0.33170 & -0.26234 & -1.55777 & 0.57086 & -0.71121 \\
\hline
JNW & $\nu = 0.1$ & 0.59869 & 0 & 0 & 0.92739 & -1.14757 & 0.14701 & -0.58785 & -0.86869 & -1.82005 & 0.47494 & -0.47888 \\
\cline{2-13}
 & $\nu = 0.5$ & 0.22527 & 0 & 0 & 0.04698 & 3.34168 & -4.35047 & 0.52143 & -0.62883 & -0.79818 & 1.41593 & -1.78545 \\
 \cline{2-13}
 & $\nu = 0.9$ & 0.06345 & 0 & 0 & -0.23279 & 2.18778 & -2.43883 & 4.30862 & -0.42314 & 4.60772 & -3.49652 & 4.07833 \\
\hline
\end{tabular}
\end{table*}
\end{center}

\subsection{Boson Stars}

Spherically symmetric solutions of the Einstein-Klein-Gordon Lagrangian
with a quadratic potential can be used to model mini boson stars (MBS;
\cite{Kaup68}). Since the matter (scalar field) constituting the MBS, in
principle, extends all the way to infinity, albeit with the scalar field
density decaying rapidly, these objects lack a sharp boundary or surface,
and also permit stable circular orbits all the way to the centre of the
spacetime. MBSs can be extremely compact with the 99\%-compactness
parameter $\mathcal{C}_{99} = M_{99}/R_{99}$ reaching values of about
0.08, where $R_{99}$ is the radius within which 99\% of the mass
($M_{99}$) is contained. This parameter is a good measure of how compact
an astrophysical object without a surface is; to compare, for a
Schwarzschild BH this value is $0.5$. Here we use the two MBS models,
denoted A and B, that were numerically obtained and studied in
\cite{Olivares+18}. These have compactnesses of $0.064$ and $0.07$, and
lie on the unstable and stable boson star branches respectively. Since
from about a few tens of Schwarzschild radii, these spacetimes look
identical to that of the Schwarzschild BH, the associated PPN parameters
$\beta$ and $\gamma$ are identical to the Schwarzschild BH values. Also,
these models lack photon spheres and regions close to the centre
contribute to the image of the boson star \cite{Olivares+18}. However, it
is discussed there that due to lower densities at the centre in these
models, a relatively dark region may be discernible.

\begin{center}
\begin{table*}
\caption{Under part I of this table, we demonstrate the efficiency of the
  current parameterisation scheme by reporting the maximum relative
  error, at fourth order, in approximating the metric functions of
  various metrics and the associated observables. Typically the relative
  error drops by more than an order of magnitude order-on-order due to
  the use of Pad\'e approximants (see table
  \ref{table:Numerical_Results_Convergence} below), and at this order
  already the typical errors are at the level of $10^{-6}$. Since this
  parametrization scheme converges rapidly with increasing order of
  approximation, those few entries that are of relatively low accuracy
  will be improved by adding a few higher-order coefficients. The
  convention we use below is that if a number is smaller than $10^{-10}$,
  we set it to zero. For brevity, we the display the order of magnitude
  within square brackets. Furthermore, since one of the objectives of
  such a parametrization scheme is to test theories of gravity, we think
  it useful to report the relative difference in the \textit{exact}
  values of important observables for a particular spacetime from the
  corresponding values for the Schwarzschild BH, under part II, i.e., we
  use the exact metric functions for part II. }
\label{table:Numerical_Results}
\renewcommand{\arraystretch}{1.3}
\begin{tabular}[t]{|c|c| |c|c|c|c|c|c| |c|c|c||}
\hline
Spacetime & Physical & \multicolumn{6}{c||}{I} & \multicolumn{3}{c||}{II} \\
\cline{3-11}
& Charge & \multicolumn{6}{c||}{Maximum relative error} & \multicolumn{3}{c||}{Exact Deviation from Schwarzschild} \\
\cline{3-11}
& & \multicolumn{6}{c||}{$|\sigma| = \left|1 - O_{\text{approx}}/O_{\text{exact}}\right|$} & \multicolumn{3}{c||}{$\delta = 1 - O_{\text{exact}}/O^{\text{Schw}}_{\text{exact}}$} \\
\cline{3-11}
& & $N^2[x)$ & $B^2[x)$ & $\xi_{\text{ps}}$ & $\Omega_{_{\text{ISCO}}}$ & $\Omega_{\text{K}}[x)$ & $\Delta \phi_{\text{GL}}[x)$ & $\xi_{\text{ps}}$ & $\Omega_{_{\text{ISCO}}}$ & $\Delta \phi_{\text{GL}}(r_{_{\text{ISCO}}})$ \\
\hline
RN & $ \bar{q} = 0.5$ & 0 & 0 & 0 & 0 & 0 & 2.37 [-7] & 4.39 [-2] & -8.21 [-2] & -6.67 [-2] \\
\cline{2-11}
& $\bar{q} = 0.9$ & 0 & 0 & 0 & 0 & 0 & 1.12 [-7] & 1.69 [-1] & -3.88 [-1] & -3.28 [-1] \\
\hline
E-ae 2 & [0.1, 0.1] & 0 & 0 & 0 & 0 & 0 & 2.95 [-8] & 4.13 [-2] & -3.60 [-2] & 1.27 [-2] \\
\cline{2-11}
$[c_{13}, c_{14}]$ & [0.9, 0.1] & 0 & 0 & 0 & 0 & 0 & 7.42 [-8] & -6.71 [-1] & 5.94 [-1] & 4.87 [-1] \\
\cline{2-11}
& [0.9, 1.7] & 0 & 0 & 0 & 0 & 0 & 1.09 [-7] & 8.39  [-1] & -4.86 [0] & 9.61 [-2] \\
\hline
E-ae 1 & $c_{13} = 0.5$ & 0 & 0 & 0 & 0 & 0 & 8.58 [-8] & -2.77 [-2] & 4.98 [-2] & 5.04 [-2] \\
\cline{2-11}
& $c_{13} = 0.9$ & 0 & 0 & 0 & 0 & 0 & 1.49 [-7] & -1.55 [-1] & 2.50 [-1] & 2.43 [-1] \\
\hline
Bardeen & $\bar{q}_{\text{m}} = 0.25$ & 0 & 0 & 0 & 1.49 [-9] & 0 & 1.20 [-7] & 1.06 [-2] & -2.16 [-2] & -2.10 [-2] \\
\cline{2-11}
& $\bar{q}_{\text{m}} = 0.75$ & 1.78 [-5] & 0 & 7.35 [-6] & 2.21 [-5] & 1.33 [-5] & 3.18 [-5] & 1.20 [-1] & -2.86 [-1] & -2.92 [-1] \\
\hline
Hayward & $\bar{l} = 0.25$ & 4.77 [-7] & 0 & 1.20 [-7] & 2.56 [-6] & 6.53 [-7] & 8.89 [-6] & 4.71 [-3] & -8.56 [-3] & -8.76 [-3] \\
\cline{2-11}
& $\bar{l} = 0.75$ & 2.88 [-4] & 0 & 1.29 [-4] & 2.26 [-4] & 2.28 [-4] & 6.86 [-4] & 5.03 [-2] & -9.25 [-2] & -9.65 [-2] \\
\hline
Bronnikov & $\bar{q}_{\text{m}} = 0.5$ & 0 & 0 & $\times$ & 0 & 0 & $\times$ & $\times$ & -8.20 [-2] & $\times$ \\
\cline{2-11}
& $\bar{q}_{\text{m}} = 1.05$ & 2.58 [-7] & 0 & $\times$ & 2.95 [-8] & 1.31 [-7] & $\times$ & $\times$ & -7.15 [-1] & $\times$ \\
\hline
EEH & [1, 0.05] & 1.74 [-5] & 0 & $\times$ & 4.44 [-7] & 9.66 [-6] & $\times$ & $\times$ & -5.90 [-1] & $\times$ \\
\cline{2-11}
$[\bar{q}_{\text{m}}, \bar{\alpha}]$ & [1, 1] & 8.69 [-5] & 0 & $\times$ & 2.22 [-4] & 7.14 [-5] & $\times$ & $\times$ & -5.76 [-1] & $\times$ \\
\hline
Frolov & [0.5, 0.25] & 1.01 [-6] & 0 & 2.66 [-7] & 4.37 [-6] & 1.19 [-6] & 2.54 [-6] & 4.99 [-2] & -9.42 [-2] & -7.92 [-2] \\
\cline{2-11}
$[\bar{q}, \bar{l}]$ & [0.5, 0.6] & 1.82 [-4] & 0 & 7.87 [-5] & 1.59 [-4] & 1.41 [-4] & 2.02 [-3] & 8.34 [-2] & -1.63 [-1] & -1.51 [-1] \\
\cline{2-11}
& [0.9, 0.25] & 3.15 [-5] & 0 & 1.47 [-5] & 3.80 [-6] & 1.94 [-5] & 4.71 [-5] & 1.83 [-1] & -4.27 [-1] & -3.70 [-1] \\
\hline
KS & $\bar{a} = 1$ & 2.74 [-7] & 0 & 6.66 [-8] & 1.45 [-6] & 3.81 [-7] & 8.68 [-7] & -7.82 [-2] & 1.23 [-1] & 9.86 [-2] \\
\cline{2-11}
& $\bar{a} = 10$ & 1.70 [-2] & 0 & 8.62 [-3] & 4.34 [-2] & 1.53 [-2] & 2.38 [-2] & -2.59 [0] & 8.81 [-1] & 7.52 [-1] \\
\hline
CFM A & $\beta = -0.9$ & 0 & 1.16 [-3] & 0 & 0 & 0 & 6.35 [-3] & 0 & 0 & 7.28 [-1] \\
\cline{2-11}
& $\beta = 0.9$ & 0 & 5.01 [-7] & 0 & 0 & 0 & 3.38 [-7] & 0 & 0 & 5.43 [-2] \\
\hline
CFM B & $\beta = 1.1$ & 0 & 5.40 [-6] & 0 & 0 & 0 & 2.27 [-6] & 0 & 0 & -5.72 [-2] \\
\cline{2-11}
& $\beta = 1.2$ & 0 & 1.87 [-3] & 0 & 0 & 0 & 4.40 [-4] & 0 & 0 & -1.18 [-1] \\
\hline
Mod. Hayward & $\bar{l} = 0.25$ & 5.03 [-7] & 2.93 [-4] & 1.20 [-7] & 2.56 [-6] & 6.53 [-7] & 2.78 [-4] & 4.71 [-3] & -8.56 [-3] & -8.74 [-3] \\
\cline{2-11}
& $\bar{l} = 0.75$ & 2.93 [-4] & 4.73 [-3] & 1.29 [-4] & 2.26 [-4] & 2.28 [-4] & 2.60 [-3] & 5.03 [-2] & -9.25 [-2] & -9.64 [-2] \\
\hline
EMd & $\bar{q} = 0.7$ & 0 & 9.69 [-9] & 0 & 0 & 0 & 7.54 [-8] & 8.67 [-2] & -1.72 [-1]  & -1.37 [-1] \\
\cline{2-11}
& $\bar{q} = 1.4$ & 1.76 [-2] & 2.21	 [-2] & 7.22 [-3] & 6.84 [-3] & 4.71 [-3] & 3.77 [-2] & 5.36 [-1] & -3.18 [0] & -3.19 [0] \\
\hline
MBS A & - & 3.85 [-4] & 2.01 [-2]  & $\times$ & $\times$ & 0 & - & $\times$ & $\times$ & $\times$ \\
\hline
MBS B & - & 5.09 [-2] & 5.95 [-2] & $\times$ & $\times$ & 0 & - & $\times$ & $\times$ & $\times$ \\
\hline
JNW & $\nu = 0.1$ & 2.10 [-3] & 4.30 [-2] & $\times$ & $\times$ & 5.64 [-3] & 1.57 [-2] & $\times$ & $\times$ & $\times$ \\
\cline{2-11}
& $\nu = 0.5$ & 2.00 [-3] & 4.54 [-3] & $\times$ & 7.81 [-3] & 9.53 [-4] & 9.21 [-4] & $\times$ & 4.14 [-1] & 1.80 [-1] \\
\cline{2-11}
& $\nu = 0.9$ & 1.40 [-3] & 3.50 [-3] & 6.95 [-4] & 1.11 [-3] & 6.65 [-4] & 2.91 [-4] & 1.18 [-2] & 1.88 [-2] & 7.72 [-3] \\
\hline
\end{tabular}
\end{table*}
\end{center}

\subsection{The Janis-Newman-Winicour Naked Singularity}

The Janis-Newman-Winicour (JNW) spacetime is also obtained as a solution
of the Einstein-Klein-Gordon Lagrangian \cite{Janis+68}, and can be
expressed more simply as in \cite{Virbhadra97}. The JNW solution, when
written in the form given in equation \eqref{eq:General_Static_Metric},
has metric functions \cite{Virbhadra97},
\begin{equation} \label{eq:JNW_Metric_Functions}
f(\rho) = g^{-1}(\rho) = \left(1 - \frac{2M}{\rho \nu}\right)^{\nu}\,,\ \
h(\rho) = \rho^2\left(1 - \frac{2M}{\rho\nu}\right)^{1 - \nu}\,,
\end{equation}
The parameter $\nu$ governs the strength of the scalar field and is given
in terms of the specific scalar charge $\bar{\Phi}$ as, $\nu = \left(1 +
\bar{\Phi}^2\right)^{-1/2}$. Clearly, depending on the strength of the
scalar field, $0 < \nu < 1$. The JNW spacetime contains a strong
curvature singularity at $\rho = 2M/\nu$, which can be seen by computing
its Kretschmann scalar $\mathcal{K} = R^{abcd}R_{abcd}$, which diverges
there\footnote{It is useful to check both the Ricci and Kretschmann
  scalars since the Ricci and Weyl scalars are known to remain finite for
  several types of solutions containing curvature singularities. For
  example, Ricci vanishes for electrovacuum solutions and Weyl vanishes
  for any conformally-flat spacetime \cite{Virbhadra+97}.}.

It can be straightforwardly seen that $h(\rho)$ is a bijective function
(in fact, it is monotonically increasing) for all $\nu$, which allows us
to recast the JNW metric into the presently desired form
\eqref{eq:General_Static_Metric_RZ}. In terms of the polar-areal radial
coordinate $r$, it can be verified that the curvature singularity is now
at $r=0$ and has zero proper area.

Since the coordinate transformation equation \eqref{eq:RZ_Gauge} for the
JNW spacetime,
\begin{equation} \label{eq:CT_JNW}
\bar{\rho}^2\left(1 - \frac{2}{\bar{\rho}\nu}\right)^{1 - \nu} = \bar{r}^2\,,
\end{equation}
is typically transcendental in nature, we solve for $\bar{\rho}(\bar{r})$
numerically. In the above, we have switched to dimensionless coordinates,
$\bar{\rho} = \rho/M$ and $\bar{r} = r/M$. For values of $\nu = 0.1, 0.3,
\cdots , 0.9$, we use a uniform grid in $\log{\bar{r}}$, with 100 points
per decade to solve equation \eqref{eq:CT_JNW}. For these values,
equation \eqref{eq:CT_JNW} reduces essentially to finding the roots of a
high-degree polynomial. The grid extends from an outer radius
$\bar{r}_{\text{max}} = 10^{7}$ down to an inner radius
$\bar{r}_{\text{min}} = 10^{-2}$. The inner grid point is sufficiently
small for our purposes since we are interested in the radial range $2
\leq \bar{r} < \infty$.

\begin{figure}
\centering
\subfigure{
\includegraphics[scale=.95]{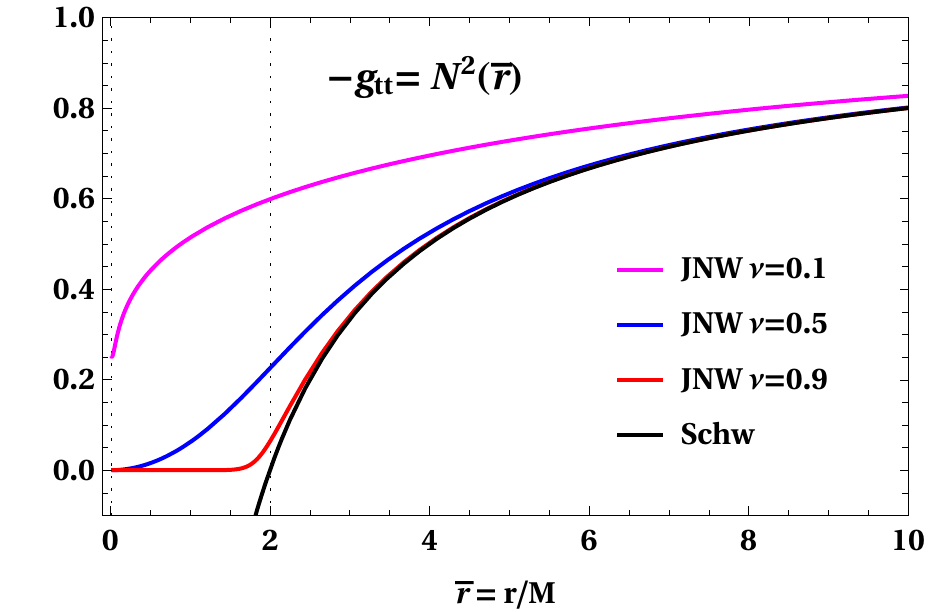}}
\subfigure{
\includegraphics[scale=.95]{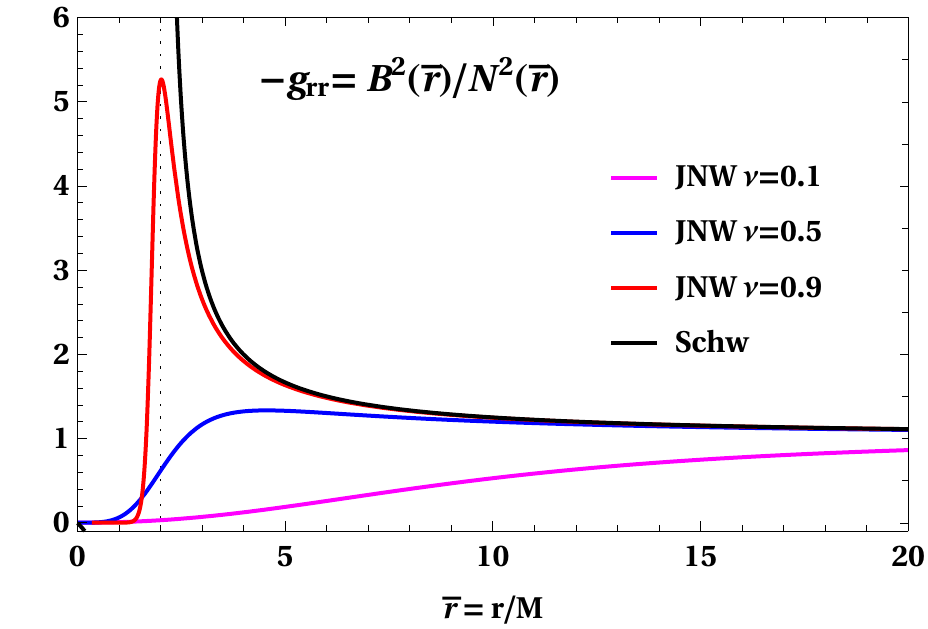}}
\caption{We show here the metric functions $g_{tt} = -N^2(\bar{r})$ and
  $g_{rr} = M^2B^2(\bar{r})/N^2(\bar{r})$ for the JNW naked singularity
  spacetime for various values of the scalar field parameter, $\nu = 0.1,
  0.5, 0.9$. The strength of the scalar field grows with decreasing
  $\nu$. We also show these metric functions for the Schwarzschild BH for
  comparison. For both spacetimes, we have set $M = 1$.}
\label{fig:JNW_Metrics}
\end{figure}

It has been shown that this spacetime contains a photon sphere for $0.5 <
\nu$ \cite{Patil_Joshi12}. For $1/\sqrt{5} \leq \nu < 0.5$, two timelike
marginally outer/inner stable circular orbits exist at
$\bar{r}_{_{\text{OSCO}}}$ and $\bar{r}_{_{\text{ISCO}}}$; stable
circular orbits extend all the way from the centre to $\bar{r} =
\bar{r}_{_{\text{OSCO}}}$ and from $\bar{r} = \bar{r}_{_{\text{ISCO}}}$
to infinity. For $0.5 \leq \nu < 1$, a single marginally stable circular
orbit remains at $\bar{r} = \bar{r}_{_{\text{ISCO}}}$ and stable circular
orbits exist only outside this location. Using the results of
\cite{Patil_Joshi12, Chowdhury+12}, we can find the locations of the
photon sphere and the timelike marginally stable circular orbits in
polar-areal radial coordinates to be,
\begin{align} \label{eq:PS_ISCO_JNW}
\bar{r}_{\text{ps}} =&\ \left(\frac{1 + 2\nu}{\nu}\right)\left(1 - \frac{2}{1 + 2\nu}\right)^{\frac{1-\nu}{2}}, \\
\bar{r}_{_{\text{ISCO/OSCO}}} =&\ \left(\frac{1 + 3\nu \pm \sqrt{5\nu^2 - 1}}{\nu}\right)\left(1 - \frac{2}{1 + 3\nu \pm \sqrt{5\nu^2 - 1}}\right)^{\frac{1-\nu}{2}}, \nonumber
\end{align}
In obtaining the above, we have simply used equation
\eqref{eq:CT_JNW}. We are able to numerically recover
$\bar{r}_{\text{ps}}$ and $\bar{r}_{_{\text{ISCO}}}$ with a relative
error of about $10^{-6}$ from the exact values reported in equation
\eqref{eq:PS_ISCO_JNW}. Note however that we are unable to obtain the
photon sphere or ISCO radius when $\bar{r}_{\text{ps}},
\bar{r}_{_{\text{ISCO}}} < \bar{r}_{\text{min}}$. This, however,
corresponds to a very small range of $0 \! < \! \nu \! < \! 1$. Since the
JNW metric has not (commonly) been reported in polar-areal coordinates,
we think it useful to display its metric functions for various values of
the scalar field parameter $\nu$ and $M=1$, in figure
\ref{fig:JNW_Metrics}.

Once we obtain $N^2(\bar{r})$ and $B^2(\bar{r})$, we make the final
change of coordinates to $x = 1 - 2/\bar{r}$ and obtain the
parametrization coefficients, which we report in table
\ref{table:eRZ_Parameters}. Finally, it can be checked that the PPN
parameters $\beta$ and $\gamma$ for this spacetime vanish identically for
all $\nu$.

\section{Discussion and Summary}
\label{sec:Discussion_Summary}

We have proposed here an extension to the RZ parametrization scheme to
allow for the characterization of arbitrary asymptotically-flat,
spherically symmetric spacetimes, including those of stars and naked
singularities. Within this scheme, we obtain highly-accurate values for
the metric functions for a variety of spacetimes: singular and
non-singular BHs from general relativity, BHs from the Einstein-aether
theory, black holes from general relativity coupled to non-linear
electrodynamics, string-inspired BH and wormhole solutions, and mini
boson stars and naked singularities in general relativity. Various other
BH solutions (and including some here) have already been studied within
this parametrization scheme and its efficiency in obtaining various
observables has been well established \cite{Kokkotas+17a, Kokkotas+17b,
  Konoplya_Zhidenko19}; see also \cite{Konoplya_Zhidenko20} and
references therein). Recently, an extension of the RZ parameterization
framework to characterize spherically symmetric BHs in higher dimensions
has also been proposed \cite{Konoplya+20}.

The shadow radii $\xi_{\text{ph}}$ of compact objects and the Kepler
orbital angular velocities $\Omega_{\text{K}}$ matter in the accretion
disks around them depend only on the $g_{tt}$-component of the
corresponding metric. Therefore accurate measurements of these
observables could be translated into constraints on the $\epsilon$ and
$a$ parameters considered here. Additionally, the profile of the
gravitational lensing angle $\Delta\phi_{\text{GL}}(r)$ for photons
emitted from the accretion disk region depends also on the
$g_{rr}$-component, and when combined with the other observables used
here, could constrain the entire metric of spherically symmetric (or
slowly-rotating) astrophysical compact objects. Other observables such as
the quasi-normal frequencies associated with a compact object also depend
on both metric functions (see equation 49 of \cite{Rezzolla_Zhidenko14}
for scalar perturbations), and combined constraints coming from all of
these observables can be simultaneously imposed in the present framework
to potentially test the underlying theories of gravity.

We have shown above that by sampling the region $0 < \epsilon, |a_i|,
|b_i|\ (0 \leq i \leq 4) < 10$, we have completely characterised all of
the BH spacetimes used in this work. This is useful when attempting to
solve the inverse problem of reconstructing a metric function
approximately given a set of observables that can essentially be
determined in terms of these variables, or equivalently as functions over
$\Pi$. Note however that these parameters may not be chosen freely. For
example, for BHs the conditions $\epsilon > -1$ and $A(x) > 0$ over $0 <
x < 1$ must always be satisfied.

If the exact relative difference in an observable $O$ for a spacetime
from its Schwarzschild BH value $O^0$ is given as $\delta = 1 - O/O^0$
and the relative error in approximating the value of $O$ is given by
$\sigma = 1 - O_{\text{approx}}/O$, then,
\begin{equation}
\delta_{\text{approx}} \equiv 1 - \frac{O_{\text{approx}}}{O^0} =  \delta + \sigma(1 - \delta)\,, 
\end{equation}
and so the absolute error in obtaining $\delta$ is,
\begin{equation} \label{eq:delta_approx_def}
\delta_{\text{approx}} - \delta =  \sigma(1 - \delta)\,.
\end{equation}
Note that $\delta$ need not be a small number; for spacetimes that
deviate significantly from the Schwarzschild BH, $\delta$ can be large
(see table \ref{table:Numerical_Results}). However, the absolute error in
obtaining $\delta$ due to approximation is clearly controlled by
$\sigma$. As we can see from table \ref{table:Numerical_Results}, where
we display both $|\sigma|$ and $\delta$, for the spacetimes considered
here, $|\sigma|$ is systematically low, about $10^{-6}$. For various
spacetimes, it is significantly lower. This means that the error in
determining whether, and how different, a particular spacetime is from
the Schwarzschild BH using EHT-observables within the present
parameterisation scheme is appreciably low.  Since this framework employs
Pad\'e approximants, the typical order-on-order decrease in $|\sigma|$ is
about $10^{-1} \! - \! 10^{-2}$, as can be seen from figure
\ref{fig:CY_vs_eRZ_Bardeen} of Appendix \ref{app:Carson_Yagi}
below. Therefore, we are able to argue comfortably that the current
framework is useful to visualise and compare various spacetimes (in terms
of the parametrization space $\Pi$ introduced above), characterise
various strong-field observables associated with them, and to enable
efficient tests of both properties of BHs from general relativity and GR
itself.

Various BH solutions considered here \cite{Bardeen68, Hayward06,
  Berglund+12, Kazakov_Solodhukin94, Casadio+02, Garcia+95,
  Gibbons_Maeda88, Garfinkle+92, Bronnikov01, Yajima_Tamaki01} were
recently studied within the same framework \cite{Rezzolla_Zhidenko14} at
first- and second-order in Pad\'e expansion
\cite{Konoplya_Zhidenko20}. It was reported there that all of these
solutions, for moderate deviations from the Schwarzschild solution, are
well approximated already at second order. While our findings are
consistent with those of \cite{Konoplya_Zhidenko20}, since the aim of the
present study is to explore the entire parameter range for these BH
solutions, and errors within this parametrization scheme typically grow
with deviation from Schwarzschild (as can be seen from table
\ref{table:Numerical_Results} above and table
\ref{table:Numerical_Results_Derivatives} in Appendix
\ref{app:Additional_Tests_eRZ} below), it becomes imperative that we
consider higher-order approximations. As has been discussed above, we
find that at the fourth-order errors in approximating metric functions
and observables are sufficiently low across the entire parameter range
for all BH solutions. Furthermore, our PPN constraint study shows that
many of the BH spacetimes considered here (Bardeen, Hayward, Modified
Hayward) satisfy the PPN constraints across their entire parameter range
(see table \ref{table:BH_Spacetimes}), and parameterizing BHs that
deviate significantly (close to their maximal deviation even) from the
Schwarzschild solution becomes important from an observational
standpoint. Also, to bring the error in approximating the deflection
angle due to gravitational lensing $\Delta\phi_{\text{GL}}(r)$ across the
entire accretion disk $r_{_{\text{ISCO}}} \! \leq \! r$ to sufficiently
low levels, we find a fourth-order approximation to be typically
necessary. A comparison between the errors reported in
\cite{Konoplya_Zhidenko20} with those reported here when approximating
the ISCO orbital angular velocity $\Omega_{_{\text{ISCO}}}$ also
demonstrates the rapidity of the convergence to the true value by going
to higher orders within the current framework, due to its use of Pad\'e
approximants. The relative error levels $|\sigma|$ reported here are
typically a few orders of magnitude smaller than the ones reported in
\cite{Konoplya_Zhidenko20}, as can be seen from table
\ref{table:Numerical_Results_Convergence} of Appendix
\ref{app:Additional_Tests_eRZ} below. For example, the errors in
approximating $\Omega_{_{\text{ISCO}}}$ for the common BH solutions vary
between $0.2-10.5\%$ and $0.04-7.95\%$ at first-and second-order
\cite{Konoplya_Zhidenko20}, while the maximum percentage error at
fourth-order is about $10^{-4}\%$ for moderate deviations from the
Schwarzschild solution. Finally, we think it useful to note that while we
have focussed on approximating observables that are associated with the
construction of the image of a compact object, a study of the
quasi-normal frequencies associated with scalar perturbations of these BH
spacetimes, which could be indicative of their gravitational wave
frequency spectrum, is also presented in \cite{Konoplya_Zhidenko20}.

We note two limitations of this framework: spacetimes that have identical
metric functions on $\bar{r}_0 \leq \bar{r} < \infty$ cannot be
distinguished between. For example, thin-shelled gravastars
\cite{Mazur_Mottola02}, whose exterior geometries are described by the
Schwarzschild metric, are hard to distinguish from a Schwarzschild black
hole in this parametrization scheme. The second limitation is that if a
metric is non-analytic, i.e., the metric functions or, as is more common,
their derivatives have discontinuities at some surface, then they cannot
be well characterized within this framework across the entire range over
which the metric is defined. Of course, the patch outside the
discontinuous surface can still be well characterized. Note that a metric
derivative discontinuity does not imply the spacetime is unphysical; this
is a common feature of various solutions that describe the collapse of
matter, and of the eventual limiting spherically symmetric spacetimes
they settle into. In these scenarios, the spacetime is divided into two
regions depending on the extent of the matter, with the interior
collapsing region matched to an appropriate exterior metric. While the
first and second fundamentals of such a spacetime (induced metric and
extrinsic curvature) are smoothly matched, the spacetime metric could
still present discontinuities on the matching surface (see for example
\cite{Shaikh+18}). In such cases, it might be possible that a two-point
or even a multi-point Pad{\'e} approximant based approach would yield
dividends (see for example Sec. 8.3 of \cite{Bender_Orszag99} for a
discussion, and for related numerical results).

Finally, we note that the low level of errors in obtaining the metric
functions up to two derivatives (see table
\ref{table:Numerical_Results_Derivatives} below) serve as a serious
impetus to attempt a study of hydrodynamics within this framework, and
potentially obtain full general-relativistic magnetohydrodynamic (GRMHD)
simulations of accretion flows around various compact objects with
state-of-the-art codes such as the Black Hole Accretion Code
(\texttt{BHAC}) \cite{Porth+17, Olivares+19}, for instance.  In fact, for
the Einstein-dilaton BH spacetime (discussed here) GRMHD simulations have
already been successfully implemented \cite{Mizuno+18}, where it has been
shown that there are clear observational differences in its image from
that of a GR Kerr BH.  Another potential application would be to study
tidal disruptions of stars and neutron stars close to compact
objects. While we do not display here the errors in obtaining the
curvature invariants $R = R^{ab}R_{ab}$ and the Kretschmann scalar
$\mathcal{K} = R^{abcd}R_{abcd}$, we find that these are also typically
approximated very well within this parametrization scheme, as can be
expected from the errors in the values of the metric and its derivatives
reported here. This implies that one can calculate the Weyl scalar
efficiently as well and potentially characterise the radii of tidal
disruption events for various spacetimes by introducing a Frenet-Serret
tetrad along static observers (see for example \cite{Kocherlakota+19} and
references therein), to provide yet another new observable to distinguish
solutions. While the spectrum of quasi-normal modes of scalar
perturbations of spacetimes within this scheme has been studied
\cite{Volkel_Kokkotas19, Konoplya_Zhidenko20, Volkel_Barausse20}, and is
somewhat representative of the spectrum of gravitational waves (GWs), a
study of the latter requires one to consider the equations of motion of
the theory of gravity that the spacetime belongs to. Since we show that
the error in approximating up to second derivatives of the metric
function across the entire exterior geometry is small already at
fourth-order in our framework, it is possible that the GW spectra of
higher-derivative gravity theories can also be obtained efficiently in
this framework.

\begin{acknowledgments}

It is a pleasure to thank Enrico Barausse, Hector Olivares, Ronak
M. Soni, and Sebastian V\"olkel for useful discussions. Support comes in
part from the ERC Synergy Grant ``BlackHoleCam: Imaging the Event Horizon
of Black Holes'' (Grant No. 610058).

\end{acknowledgments}

\begin{appendix}

\section{Tortoise Coordinate for Black-Hole Spacetimes}
\label{app:Other_Useful_Coordinates}

If the metric $\bm{g}$ of a spacetime can be brought into the form
$\bm{g}(\bm{x}) = \mathscr{C}(\bm{x})\bm{\eta}$, then such
a metric $\bm{g}$ is conformally flat and the coordinates $x^\mu$
are called conformally flat coordinates; here $\bm{\eta}$ is the
Minkowski metric tensor. Such coordinates are particularly useful when
attempting to study the global causal structure of a spacetime. Radial
null geodesics in the associated spacetime diagrams are given by
$45^\circ$ lines, similar to flat-space spacetime diagrams.

Since all 2D geometries are conformally flat, we can find them for the $t
\! - \! r$ plane of arbitrary BH ($n_0 = 0$) spherically symmetric metrics
\eqref{eq:General_Static_Metric_RZ} as,
\begin{equation} 
\text{d}s^2_{d\theta = d\phi = 0} = - N^2(r)\left(\text{d}t^2 + \frac{B^2(r)}{N^4(r)}\text{d}r^2\right) =  - N^2(r)\left(\text{d}t^2 + \text{d}r_*^2\right)\,, \nonumber
\end{equation}
where the equation to achieve the coordinate tranformation $r \rightarrow r_*$ can be read off from above as,
\begin{equation}
\frac{dr_*}{dr} = \frac{B(r)}{N^2(r)}\,.
\end{equation}
In terms of the function $A$ defined in equation \eqref{eq:Ax_eRZ} above, this is simply,
\begin{equation}
\frac{dr_*}{dr} = \left(1-  \frac{r_0}{r}\right)^{-1}\frac{B(r)}{A(r)}\,.
\end{equation}
For the Schwarzschild spacetime, since $A(r) = B(r) = 1$, this coordinate $-\infty < r_* < \infty$ is exactly the familiar Tortoise coordinate,
\begin{equation}
r_* = r + r_0\ln{\left|1 - \frac{r_0}{r}\right|}\,.
\end{equation}
We can now relate the two conformal coordinates, $x$ and $r_*$, for a
generic BH spacetime via,
\begin{equation}
\frac{dr_*}{dx} = \frac{r_0}{x(1-x)^2}\frac{B(x)}{A(x)}\,.
\end{equation}

\begin{center} 
\begin{figure}[h!]
\includegraphics[scale=.65]{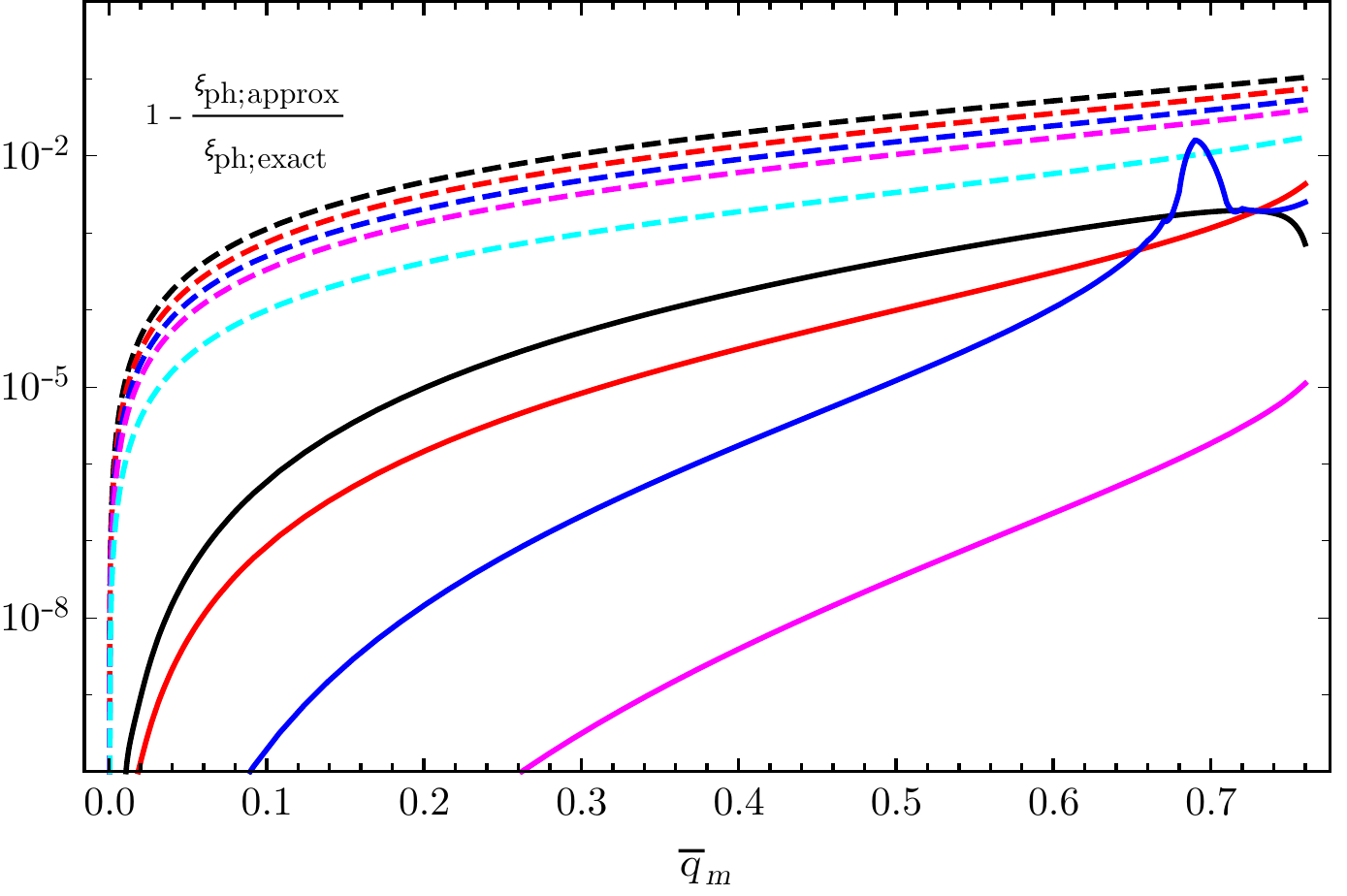}
\caption{We consider here the Bardeen BH \cite{Bardeen68}, with specific
  magnetic charge $0 < \bar{q}_{\text{m}} \lesssim 0.77$, and show the
  absolute relative error in estimating the exact value of its unstable
  circular photon-orbit impact parameter $\xi_{\text{ps;exact}}$ within
  the Carson-Yagi parametrization scheme (in dashed lines), which uses
  Taylor expansions to characterize metric functions, and in our scheme,
  which uses expansions of Pad\'e approximants. Considering $\alpha_{13}$
  to be the first non-trivial parameter of the CY approximation scheme,
  lines in the same color correspond to the same number of approximation
  parameters. To be clear, dashed-black and solid-black correspond to the
  $\alpha_{1i>13} = 0$ Carson-Yagi and $a_{i>2}=0$ Pad\'e approximation
  respectively. We also show in dashed-cyan the seventh-order CY relative
  error, i.e., for $\alpha_{1i>19} = 0$. It is apparent that the Pad\'e
  approximation does better already at first-order relative to the
  seventh-order Taylor expansion. Note also that the rapidity of
  convergence is much higher when employing Pad\'e approximants.}
\label{fig:CY_vs_eRZ_Bardeen}
\end{figure}
\end{center}

\section{Comaprison with the Carson-Yagi Parametrisation Scheme}
\label{app:Carson_Yagi}

The metric of a spherically symmetric ($S = 0$) BH spacetime in the
Carson-Yagi (CY) parametrization scheme \cite{Carson_Yagi20} can be
expressed as,
\begin{equation}
\text{d}s^2 =  -\left(1 - \frac{2M}{r}\right)A_1^{-2}(r)~\text{d}t^2 + \left(1 - \frac{2M}{r}\right)^{-1}A_5^{-1}(r)~\text{d}r^2 + r^2\text{d}\Omega_2^2\,,
\end{equation}
where the functions $A_1(r)$ and $A_5(r)$ measure the deviation of an
arbitrary BH metric from Schwarzschild. An asymptotic Taylor expansion for
these functions is then employed to characterize them as,
\begin{equation}
A_i(r) = 1 + \sum_{n=1}^{\infty}\alpha_{i0}\left(\frac{M}{r}\right)^n\,. 
\end{equation}
To compare the rapidity of convergence of this scheme against the one
used in the current work, we study the Bardeen-BH spacetime
\cite{Bardeen68}. For this BH, we obtain the CY functions, $A_1$ and
$A_5$, from Table \ref{table:BH_Spacetimes} to be,
\begin{align} \label{eq:A1_A5_Bardeen}
A_1(\bar{r}) =&\ \left(1 - \frac{2}{\bar{r}}\right)^{1/2}\left(1 - \frac{2\bar{r}^2}{\left(\bar{r}^2 + \bar{q}_{\text{m}}^2\right)^{3/2}}\right)^{-1/2}\,, \\
A_5(\bar{r}) =&\ \left(1 - \frac{2}{\bar{r}}\right)^{-1}\left(1 - \frac{2\bar{r}^2}{\left(\bar{r}^2 + \bar{q}_{\text{m}}^2\right)^{3/2}}\right)\,,
\end{align}
where in the above we have introduced $\bar{r} = r/M$ for brevity. Also,
the parameter $\bar{q}_{\text{m}}$ corresponds to the parameter $g/M$ of
\cite{Carson_Yagi20}. We obtain now the CY coefficients $\alpha_{i0}$ up
to fifth order for this spacetime as,
\begin{align}
\alpha_{10} =&\ 1,\ \alpha_{13} = -\frac{3\bar{q}_{\text{m}}^2}{2},\ \alpha_{14} = -3\bar{q}_{\text{m}}^2,\ \alpha_{15} = -6\bar{q}_{\text{m}}^2 + \frac{15\bar{q}_{\text{m}}^4}{8}\,, \nonumber \\
\alpha_{50} =&\ 1,\ \alpha_{53} = 3\bar{q}_{\text{m}}^2,\ \alpha_{54} = 6\bar{q}_{\text{m}}^2,\ \alpha_{55} = 12\bar{q}_{\text{m}}^2 - \frac{15\bar{q}_{\text{m}}^4}{8}\,,
\end{align}
which can be verified to match the expressions in table 1 of
\cite{Carson_Yagi20}. From the form of the functions in equation
\eqref{eq:A1_A5_Bardeen}, it is clear that for appreciable specific
magnetic charges $0 \lnsim \bar{q}_{\text{m}}$, close to the horizon,
which lies between $1.23 \lesssim \bar{r}_{\text{H}} \lesssim 1.99$,
arbitrarily high-order coefficients might become important in this
parametrization scheme. We demonstrate this by showing in figure
\ref{fig:CY_vs_eRZ_Bardeen} how the impact parameter of a photon on the
unstable circular geodesic in this spacetime, which is a near-horizon
observable, is approximated within both the CY parametrization scheme and
in our scheme.

\begin{center}
\begin{table}
\caption{Here we show the relative error in approximating the first and
  second derivatives of the metric functions for all of the BH and non-BH
  spacetimes considered here.}
\label{table:Numerical_Results_Derivatives}
\begin{tabular}[t]{||c|c||c|c||c|c||}
\hline
Spacetime & Physical & \multicolumn{4}{c||}{Maximum relative error} \\
\cline{3-6}
& Charge & $dN/dx$ & $d^2N/dx^2$ & $dB/dx$ & $d^2B/dx^2$ \\
\hline
RN & $ \bar{q} = 0.5$ & 0 & 0 & 0 & 0 \\
\cline{2-6}
& $\bar{q} = 0.9$ & 0 & 0 & 0 & 0 \\
\hline
E-ae 2 & [0.1, 0.1] & 0 & 0 & 0 & 0 \\
\cline{2-6}
[$c_{13}, c_{14}$] & [0.1, 0.9] & 0 & 0 & 0 & 0 \\
\cline{2-6}
& [0.9, 1.7] & 0 & 0 & 0 & 0 \\
\hline
E-ae 1 & $c_{13} = 0.5$ & 0 & 0 & 0 & 0 \\
\cline{2-6}
& $c_{13} = 0.9$ & 0 & 0 & 0 & 0 \\
\hline
Bardeen & $\bar{q}_{\text{m}} = 0.25$ & 0 & 3.04 [-9] & 0 & 0 \\
\cline{2-6}
& $\bar{q}_{\text{m}} = 0.75$ & 2.10 [-5] & 1.09 [-3] & 0 & 0 \\
\hline
Hayward & $\bar{l} = 0.25$ & 8.25 [-7] & 5.35 [-6] & 0 & 0 \\
\cline{2-6}
& $\bar{l} = 0.75$ & 3.39 [-4] & 5.43 [-3] & 0 & 0 \\
\hline
Bronnikov & $\bar{q}_{\text{m}} = 0.5$ & 0 & 0 & 0 & 0 \\
\cline{2-6}
& $\bar{q}_{\text{m}} = 1.05$ & 2.74 [-7] & 7.17 [-6] & 0 & 0 \\
\hline
EEH & [1, 0.05] & 1.96 [-5] & 3.93 [-2] & 0 & 0 \\
\cline{2-6}
[$\bar{q}_{\text{m}}, \bar{\alpha}$] & [1, 1] & 1.31 [-4] & 4.75 [-2] & 0 & 0 \\
\hline
Frolov & [0.5, 0.25] & 1.60 [-6] & 1.22 [-5] & 0 & 0 \\
\cline{2-6}
[$\bar{q}, \bar{l}$] & [0.5, 0.6] & 2.16 [-4] & 3.76 [-3] & 0 & 0 \\
\cline{2-6}
& [0.9, 0.25] & 3.59 [-5] & 9.83 [-4] & 0 & 0 \\
\hline
KS & $\bar{a} = 1$ & 4.60 [-7] & 2.93 [-6] & 0 & 0 \\
\cline{2-6}
& $\bar{a} = 10$ & 3.80 [-2] & 6.93 [-2] & 0 & 0 \\
\hline
CFM A & $\beta = -0.9$ & 0 & 0 & 2.03 [-3] & 2.01 [-2] \\
\cline{2-6}
& $\beta = 0.9$ & 0 & 0 & 2.65 [-6] & 1.74 [-5] \\
\hline
CFM B & $\beta = 1.1$ & 0 & 0 & 4.46 [-5] & 7.7 [-5] \\
\cline{2-6}
& $\beta = 1.2$ & 0 & 0 & 7.26 [-3] & 4.04 [-3] \\
\hline
Mod. Hayward & $\bar{l} = 0.25$ & 8.25 [-7] & 5.35 [-6] & 9.96 [-2] & 1.94 [-1] \\
\cline{2-6}
& $\bar{l} = 0.75$ & 3.39 [-4] & 5.43 [-3] & 9.20 [-2] & 1.69 [-1] \\
\hline
EMd & $\bar{q} = 0.7$ & 0 & 0 & 1.43 [-6] & 1.32 [-6] \\
\cline{2-6}
& $\bar{q} = 1.4$ & 1.74 [-2] & 4.29 [-2] & 3.09 [-2] & 6.58 [-2] \\
\hline
MBS A & - & 4.79 [-2] & 1.92 [-2] & 8.06 [-2] & 3.68 [-2] \\
\hline
MBS B & - & 1.88 [-1] & 2.94 [-1] & 2.22 [-1] & 5.08 [-1] \\
\hline
JNW & $\nu = 0.1$ & 1.45 [-2] & 3.55 [-2] & 4.04 [-1] & 2.60 [-2] \\
\cline{2-6}
& $\nu = 0.5$ & 4.31 [-3] & 2.51 [-1] & 1.02 [-2] & 2.91 [-3] \\
\cline{2-6}
& $\nu = 0.9$ & 1.61 [-3] & 1.31 [-2] & 5.27 [-2] & 2.63 [-2] \\
\hline
\end{tabular}
\end{table}
\end{center}

\begin{center}
\begin{table}
\caption{Here we demonstrate the rapidity of convergence of the
  parametrization scheme used here by estimating the relative error in
  obtaining the unstable circular null geodesic impact factor
  $\xi_{\text{ps}}$ and ISCO frequency $\Omega_{_{\text{ISCO}}}$ at
  order-two and -four in the current parametrization scheme. We have
  omitted below those spacetimes for which the only relevant metric
  function $N$ for these observables is already exactly recovered at a
  lower order.}
\label{table:Numerical_Results_Convergence}
\begin{tabular}[t]{||c|c||c|c||c|c||}
\hline
Spacetime & Physical & \multicolumn{4}{c||}{Maximum relative error} \\
\cline{3-6}
& Charge & \multicolumn{2}{c||}{$\xi_{\text{ps}}$} & \multicolumn{2}{c||}{$\Omega_{_{\text{ISCO}}}$} \\
\cline{3-6}
& & $a_3=0$ & $a_5=0$ & $a_3=0$ & $a_5=0$ \\
\hline
Bardeen & $\bar{q}_{\text{m}} = 0.25$ & 3.71 [-6] & 0 & 2.00 [-6] & 1.49 [-9] \\
\cline{2-6}
& $\bar{q}_{\text{m}} = 0.75$ & 3.10 [-3] & 7.35 [-6] & 6.74 [-3] & 2.21 [-5] \\
\hline
Hayward & $\bar{l} = 0.25$ & 1.78 [-5] & 1.20 [-7] & 3.28 [-6] & 2.56 [-6] \\
\cline{2-6}
& $\bar{l} = 0.75$ & 5.44 [-2] & 1.29 [-4] & 1.02 [-1] & 2.26 [-4] \\
\hline
Bronnikov & $\bar{q}_{\text{m}} = 0.5$ & $\times$ & $\times$ & 0 & 0 \\
\cline{2-6}
& $\bar{q}_{\text{m}} = 1.05$ & $\times$ & $\times$ & 1.22 [-3] & 2.95 [-8] \\
\hline
EEH & [1, 0.05] & $\times$ & $\times$ & 8.31 [-4] & 4.44 [-7] \\
\cline{2-6}
[$\bar{q}_{\text{m}}, \bar{\alpha}$] & [1, 1] & $\times$ & $\times$ & 4.68 [-3] & 2.22 [-4] \\
\hline
Frolov & [0.5, 0.25] & 4.18 [-5] & 2.66 [-7] & 2.09 [-5] & 4.37 [-6] \\
\cline{2-6}
[$\bar{q}, \bar{l}$] & [0.5, 0.6] & 1.17 [-1] & 7.87 [-5] & 5.66 [-1] & 1.59 [-4] \\
\cline{2-6}
& [0.9, 0.25] & 1.10 [-2] & 1.47 [-5] & 2.12 [-1] & 3.80 [-6] \\
\hline
KS & $\bar{a} = 1$ & 9.31 [-6] & 6.66 [-8] & 1.39 [-6] & 1.45 [-6] \\
\cline{2-6}
& $\bar{a} = 10$ & 2.78 [-2] & 8.62 [-3] & 1.11 [-1] & 4.34 [-2] \\
\hline
Mod. Hayward & $\bar{l} = 0.25$ & 5.44 [-2] & 1.29 [-4] & 1.02 [-1] & 2.26 [-4] \\
\cline{2-6}
& $\bar{l} = 0.75$ & 1.78 [-5] & 1.20 [-7] & 3.28 [-6] & 2.56 [-6] \\
\hline
EMd & $\bar{q} = 0.7$ & 4.43 [-7] & 0 & 3.97 [-7] & 0 \\
\cline{2-6}
& $\bar{q} = 1.4$ & 7.62 [-1] & 7.22 [-3] & 2.93 [-1] & 6.84 [-3] \\
\hline
MBS A & - & $\times$  & $\times$  & $\times$  & $\times$  \\
\hline
MBS B & - & $\times$ & $\times$ & $\times$ & $\times$ \\
\hline
JNW & $\nu = 0.1$ & $\times$  & $\times$ & $\times$ & $\times$ \\
\cline{2-6}
& $\nu = 0.5$ & $\times$  & $\times$ & 3.54 [-3] & 7.81 [-3] \\
\cline{2-6}
& $\nu = 0.9$ & 2.86 [-3] & 6.95 [-4] & 7.85 [-3] & 1.11 [-3] \\
\hline
\end{tabular}
\end{table}
\end{center}

\section{Goodness of Approximation and Convergence Tests}
\label{app:Additional_Tests_eRZ}

The fourth-order approximations of the metric functions for the various
spacetimes under consideration here are smooth throughout the range of
the conformal coordinate $0 \!  \leq x \! < \! 1$. To demonstrate the
accuracy in obtaining the metric function up to two derivatives, we
report the relative errors in approximating them in table
\ref{table:Numerical_Results_Derivatives}, which are typically
appreciably low.

We also demonstrate the rapidity of the order-on-order convergence of our
parametrization scheme by estimating the relative error in obtaining the
unstable circular null geodesic impact factor $\xi_{\text{ps}}$ and ISCO
frequency $\Omega_{_{\text{ISCO}}}$ at the second and fourth order in
Pad{\'e} expansion. This is reported in table
\ref{table:Numerical_Results_Convergence}. We gain about two orders of
accuracy by going two orders higher in the Pad{\'e} expansion.

\end{appendix}
\end{document}